\documentclass{iopart}

\usepackage{fleqn}

\usepackage{ifthen}
\usepackage{ifpdf}

\usepackage{latexsym}
\usepackage{txfonts}
\usepackage{amssymb}
\usepackage{bm}

\ifpdf
\usepackage{graphicx}
\usepackage{epstopdf}
\else
\usepackage{graphicx}
\usepackage{epsfig}
\fi

\newcommand{\const}{\mbox{const}}
\newcommand{\trc}{\mbox{trace}}

\newcommand{\im}{\mbox{Im}}

\newcommand{\eexp}{\mbox{e}^}

\newcommand{\tbox}[1]{\mbox{\tiny #1}}

\newcommand{\amatrix}[1]{\matrix{#1}}

\newcommand{\be}[1]{\begin{eqnarray}\ifthenelse{#1=-1}{\nonumber}{\ifthenelse{#1=0}{}{\label{e#1}}}}
\newcommand{\ee}{\end{eqnarray}}

\newcommand{\hide}[1]{ }
\newcommand{\Cn}[1]{\begin{center} #1 \end{center}}
\newcommand{\mpg}[2][\hsize]{\begin{minipage}[b]{#1}{#2}\end{minipage}}
\newcommand{\putgraph}[2][\hsize]   {\includegraphics[#1]{#2} }

\begin{document}


\hide{

LIST OF FIGURES

-- Fig1 (models)
pmp_scat
pmt_fig_d
ClosedGeometry2Deltas
ClosedGeometry2DeltasWithPerturbation

-- Fig2 (models/schematic)
ClosedGeometry
OpenGeometry3

-- Fig.3 (cycles)
PumpingCycle2
pmp_cyc_b3

-- Fig.4 (cycles/B)
pmp_cyc_a4
pmp_cyc_a3

-- Fig.5  (Q, classical) 
ClassicalChargeAllBW
ClassicalChargelogX2BW

-- Fig.6 (sigma)
DensityOfDegeneracies_normalX2allBW

-- Fig.7 (Q, quantum)
OneDegeneracyRoutesDiagram
QuantumChargeDiagram
NDegeneraciesRoutesDiagram
QuantumChargeNDegeneraciesDiagram

-- Fig.8 (degeneracies, 2delta)
Diagram7LevelsDegeneraciesFlat
Diagram7LevelsDegeneraciesAndAvoidedCrossings2

-- Fig.9 (degeneracies, 3delta)
Diagram7LevelsDegeneraciesPerturbation

-- Fig.10 (Q, numerical)
QuantumChargeRoutesBW
QuantumChargeBW

-- Fig.11 (g1, chaotic)
OneDeltaOneComplexBarrierGraphBW

-- Fig.12 (UCF)
DensityLowTransmissionLogBW

}


\title[Quantum Stirring]
{Quantum stirring of particles in closed devices}

\author{Gilad Rosenberg and Doron Cohen}

\address{
Department of Physics, Ben-Gurion University, Beer-Sheva 84105, Israel
}


\begin{abstract}
We study the quantum analog of stirring
of water inside a cup using a spoon.
This can be regarded as a prototype example
for quantum pumping in closed devices.
The current in the device is induced by
translating a scatterer. Its calculation is done using the Kubo formula
approach.
The transported charge is expressed as a line integral that encircles chains
of Dirac monopoles. For simple systems
the results turn out to be counter intuitive:
e.g. as we move a small scatterer ``forward"
the current is induced ``backwards".
One should realize that the route towards
quantum-classical correspondence has to
do with ``quantum chaos" considerations,
and hence assumes greater complexity of the device. We also point out the
relation to the
familiar $S$ matrix formalism which is used
to analyze quantum pumping in open geometries.
\end{abstract}

\section{Introduction}

Consider a closed ring that contains particles (Fig.1a). 
Assume that one wants to create a current in this ring. 
If the particles are charged then one way to do it 
is by creating an electro motive force (EMF). 
This can be induced by varying an Aharonov-Bohm flux $\Phi$, 
such that by Faraday's law $\mbox{EMF}=-\dot{\Phi}$.
But there is another way to create a current 
that does not involve EMF, and hence does not assume 
charged particles. The idea is to change 
in time the scalar potential $V(\bm{r}; X_1(t),X_2(t))$.  
Here $\bm{r}$ is the coordinate of a representative particle 
in the ring, while $X_1$ and $X_2$ are some control 
parameters. By making a cycle in the $(X_1,X_2)$ space 
we can push non-zero net charge $Q$ through the system. 
Thus an ``AC driving" gives rise to a ``DC" component 
in the current.  This is known in the literature as ``quantum pumping". 

\vspace{3mm}

\mpg{
\Cn{
\putgraph[height=0.2\hsize]{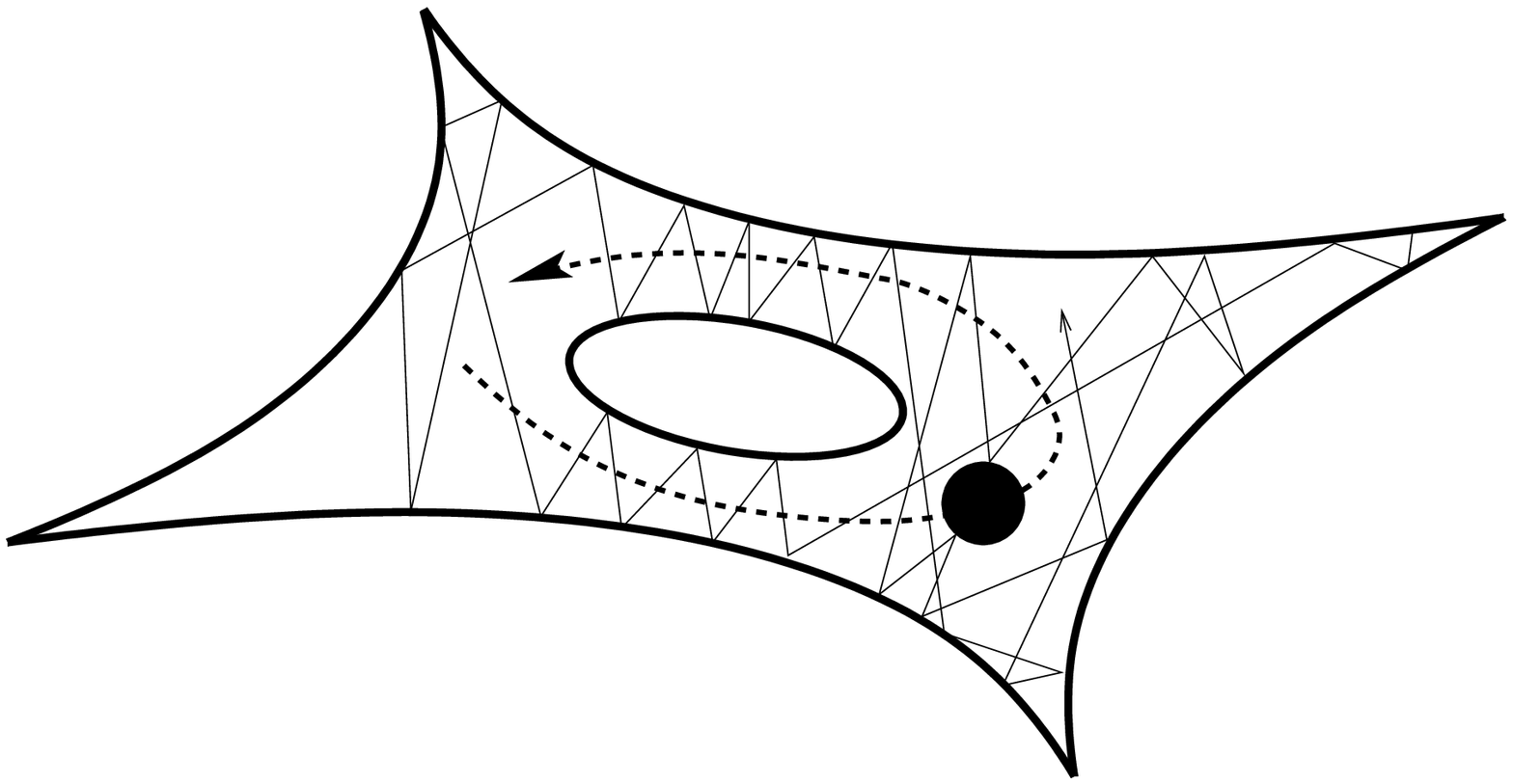}
\\ 
\putgraph[height=0.2\hsize]{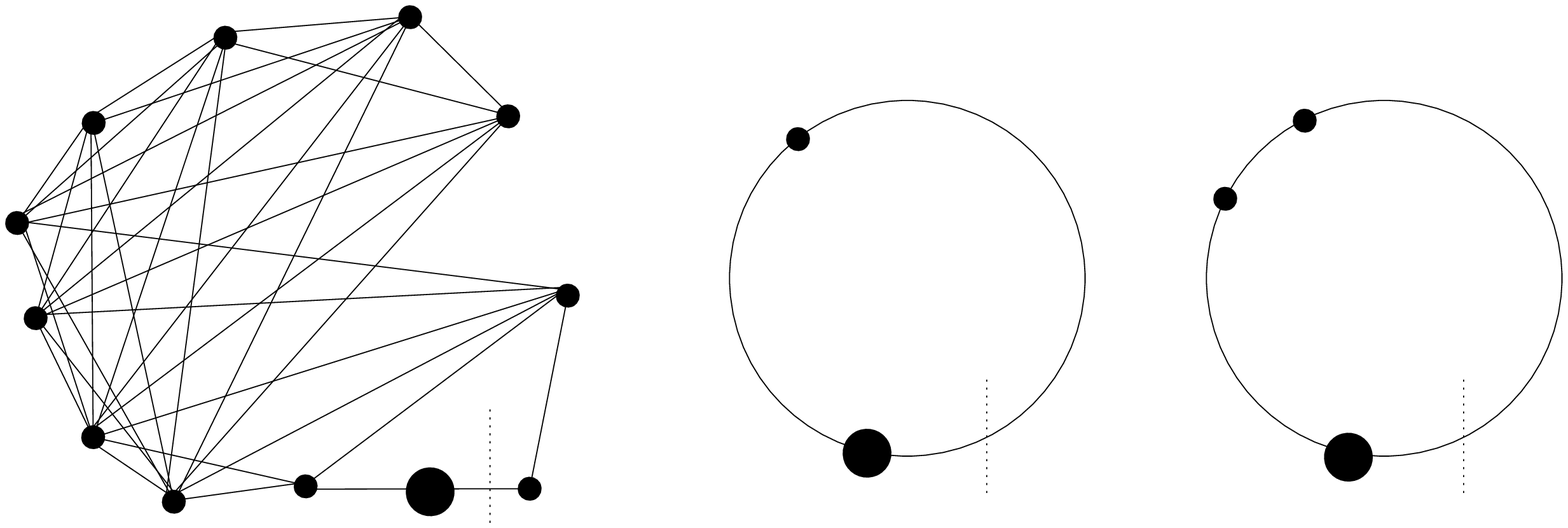}
}

{\footnotesize 
{\bf Fig.1.} 
Models for the analysis of quantum stirring. 
(a) Upper panel: A scatterer (big black dot) 
is translated inside a Sinai billiard. 
A chaotic trajectory of a representative 
particle in this billiard is illustrated. 
(b) Lower panels: Network models for 
quantum stirring. The scatterer 
(big black dot) is translated along 
one of the bonds. The vertical 
dotted line is the section through 
which the current is measured. 
From left to right: chaotic network;  
double barrier model; triple barrier model. 
}
}

\vspace{3mm}

\mpg{
\putgraph[width=0.4\hsize]{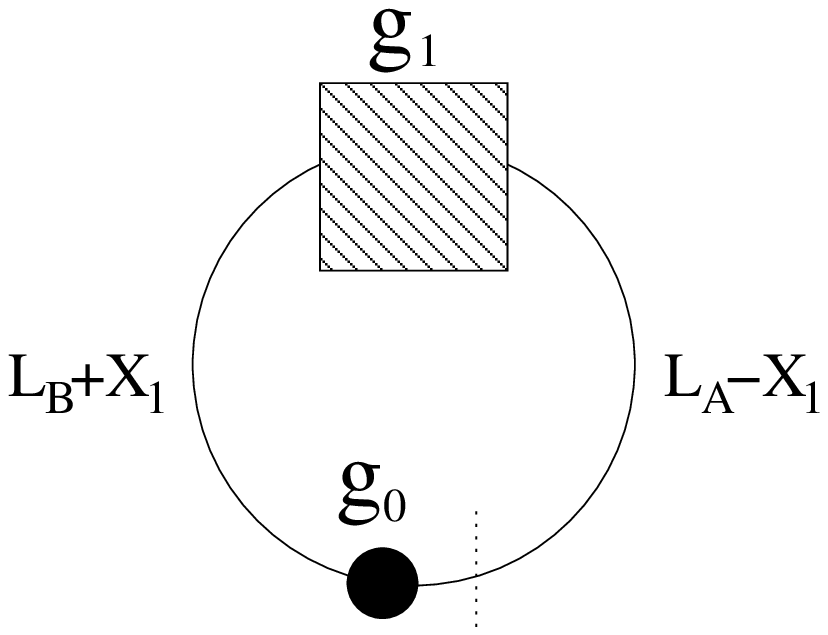}
\hspace{0.1\hsize}
\putgraph[width=0.4\hsize]{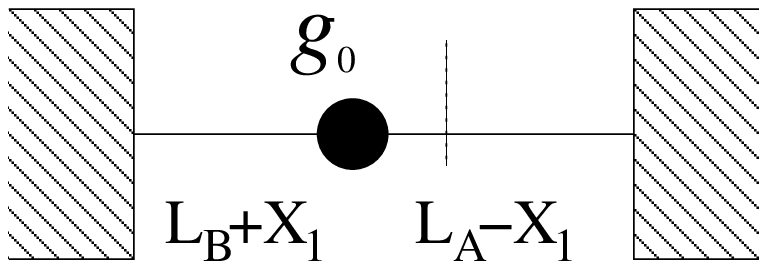} 

{\footnotesize 
{\bf Fig.2.} 
(a) Left panel: A schematic representation  
of the network model. The vertical 
dotted line is the section through 
which the current is measured.  
The moving scatterer is indicated by its transmission $g_0$,   
while $X_1$ is its displacement along the bond. 
(b) Right panel: The corresponding 
open geometry where the left and right 
leads are connected to reservoirs 
with the same chemical potentials.
}
}

\vspace{3mm}

In this paper we would like to consider a prototype 
pumping problem, which we call ``quantum stirring". 
It is the simplest scheme to create a current 
with a non-vanishing DC component. Referring to Fig.2   
we define $X_1$ as the location of a scatterer,  
while $X_2$ is its ``size". By ``size" we mean either 
the cross section or the reflection coefficient.
One can regard the scatterer as a ``piston" 
or as a ``spoon" with which it is possible 
to ``push" the particles.  A prototype example 
for a pumping cycle is illustrated in Fig.3. 
During the main stage of the cycle the 
scatterer is translated to the right 
a distance $\Delta X_1$. 
Consequently a charge $Q$ is transported.    
In the second stage the size of the scatterer
is ``lowered", and it is displaced back to 
its original location, where its original ``size" 
is restored. By repeating this cycle many times 
we can create a current with a DC component.

In the following analysis we assume that the system 
consists of non-interacting spinless particles.  
All the particles have (formally) charge~$e$, 
even if they are not actually charged. 
We assume that there is no magnetic field in the system. 
Still, for the sake of a later mathematical 
formulation, it is convenient to introduce 
a third parameter $X_3=\Phi$, 
where $\Phi$ is an Aharonv-Bohm flux. 
The pumping cycle in the ${(X_1, X_2, X_3)}$ space 
is illustrated in Fig.3.

\vspace{3mm}

\mpg{
\Cn{
\putgraph[width=0.45\hsize]{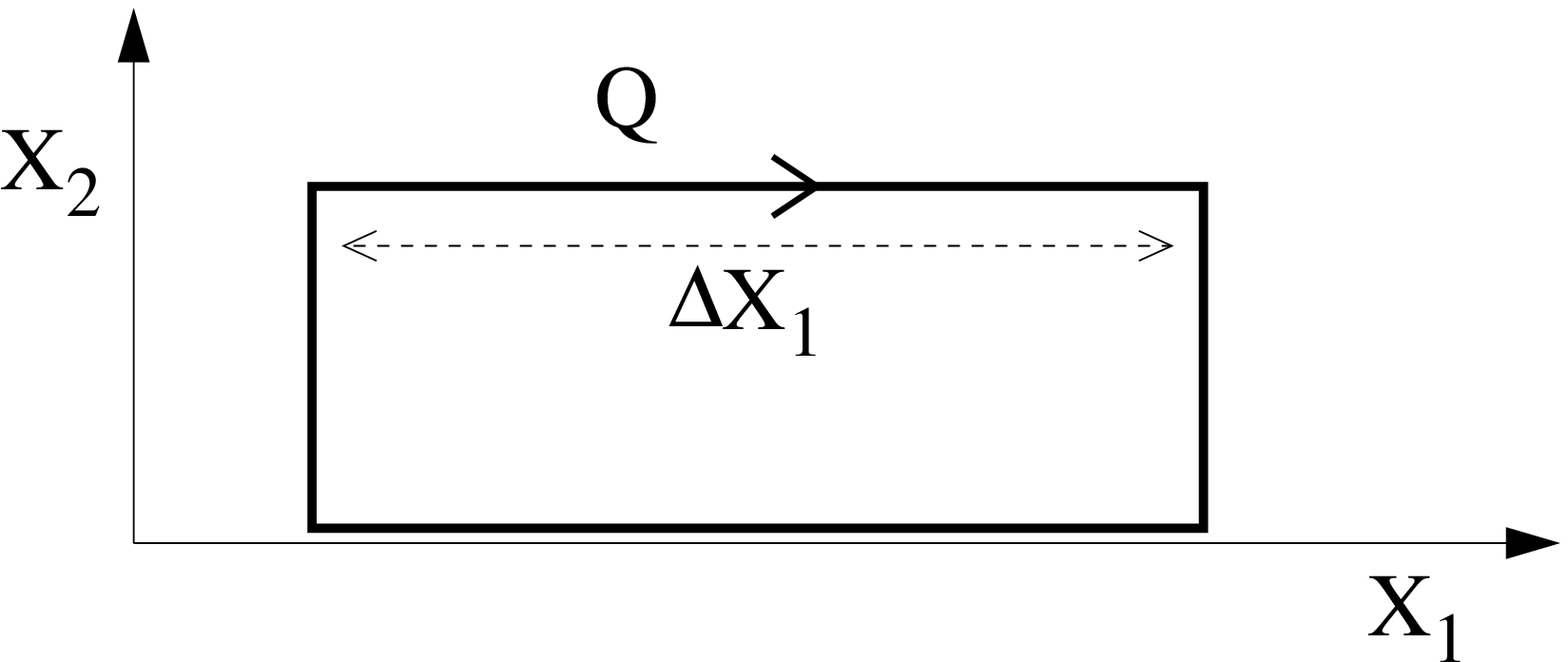}
\putgraph[width=0.4\hsize]{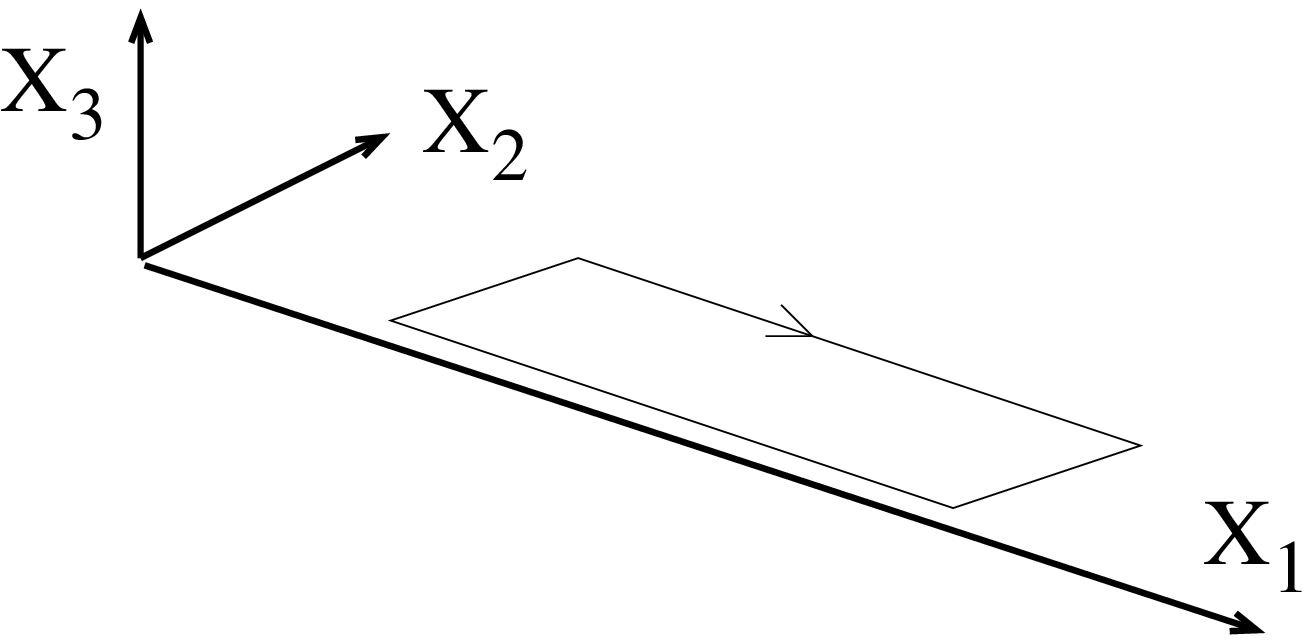}
}

{\footnotesize 
{\bf Fig.3.} 
A prototype example for a pumping cycle. 
During the main stage of the cycle the 
scatterer is translated to the right 
a distance $\Delta X_1$. Consequently 
a charge $Q$ is transported.    
(a) Left panel:  
The pumping cycle in the 2-dimensional $(X_1,X_2)$ plane. 
(b) Right panel: The same pumping cycle in 
the three dimensional ${(X_1, X_2, X_3)}$ space, 
where $X_3=\Phi$ is the Aharonov Bohm flux 
via the ring. 
}
}

\vspace{3mm}

\mpg{
\Cn{
\putgraph[width=0.3\hsize]{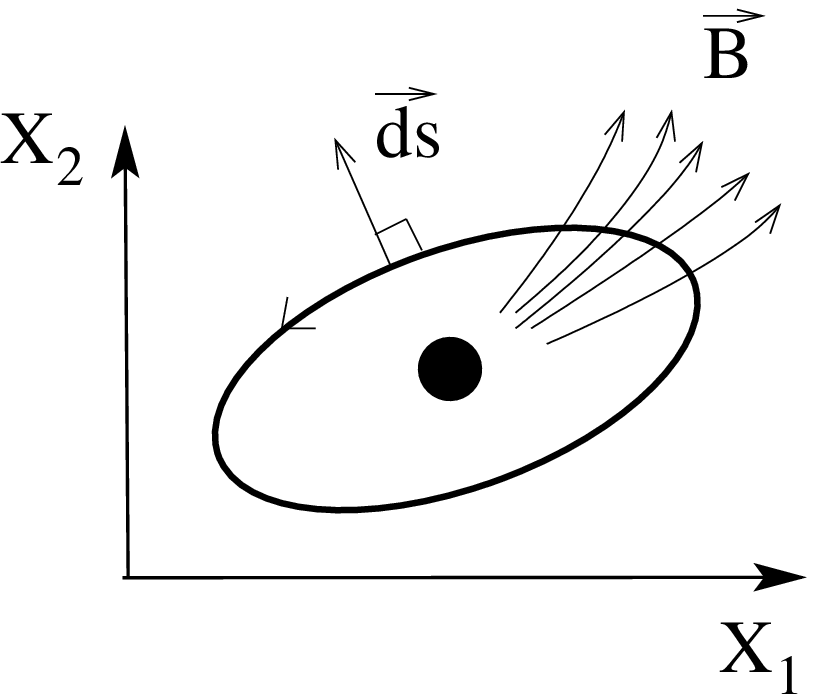}
\hspace{0.1\hsize}
\putgraph[width=0.3\hsize]{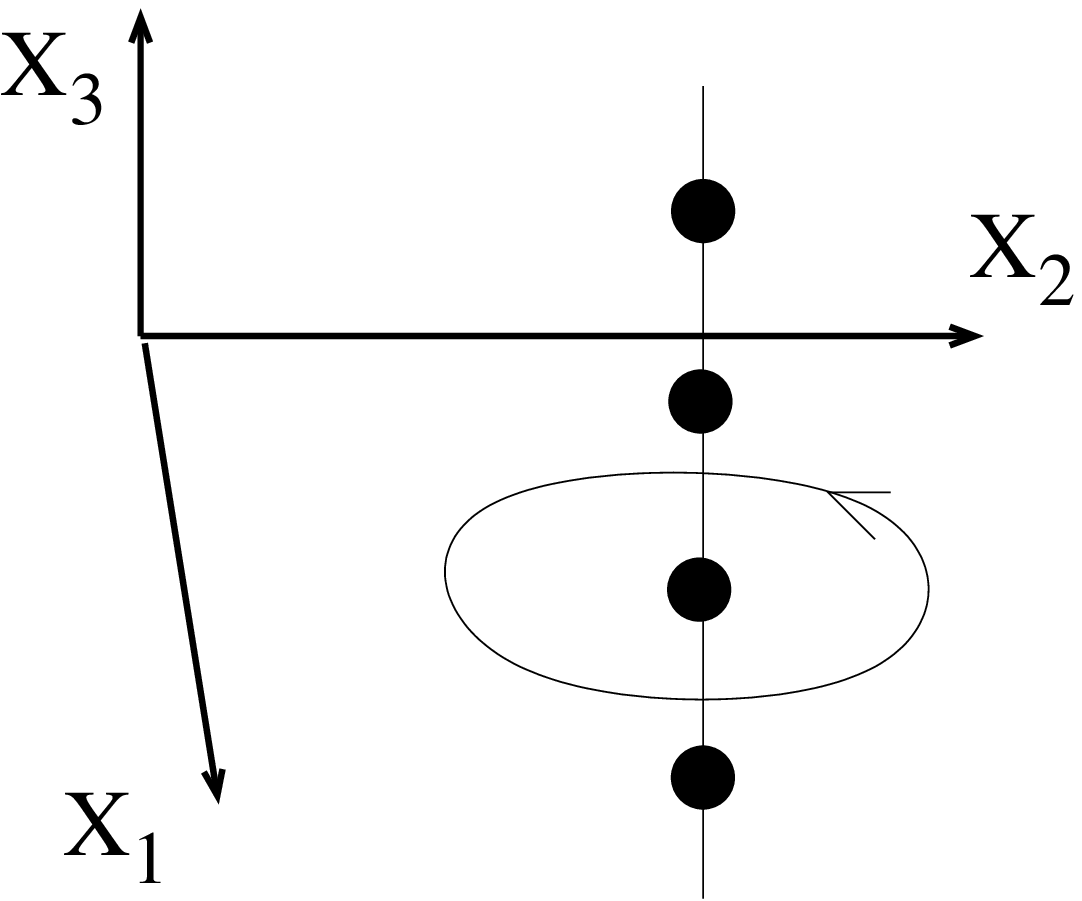}
}

{\footnotesize 
{\bf Fig.4.}
(a) Left panel: The calculation of the charge $Q$ is a line integral 
over $G$ that can be regarded as a calculation 
of the flux of $\bm{B}$ via a two dimensional curve.  
$\vec{ds}$ is a normal vector to the pumping cycle.
The black dot in the middle symbolizes the 
presence of ``magnetic charge" which is characterized 
by a density $\sigma(X_1,X_2)$. In the quantum mechanical 
analysis this should be understood as the density 
of ``Dirac chains". 
(b) Right panel: In the embedding ${(X_1, X_2, X_3)}$ space 
the magnetic charge is organized as vertical charged chains.
Each chain consists of ``Dirac monopoles" which are located 
at $\vec{X}$ points where an occupied level has a degeneracy 
with a nearby level. The ellipse represents a possible 
pumping cycle that may encircle either one or many chains. 
}
}

\vspace{3mm}

\subsection{Linear response theory and the Dirac chains picture} 

We are going to analyze the stirring problem 
within the framework of linear response theory. 
If we have EMF then we expect to get in the DC limit 
Ohm law $\mathcal{I}=-G\dot{\Phi}$, 
while if we change slowly either $X_1$ or $X_2$ 
we expect to get in the DC limit  $\mathcal{I}=-G^1\dot{X}_1$ 
or  $\mathcal{I}=-G^2\dot{X}_2$ respectively. 
So in general we can write 
\be{1}
Q = \oint_{\tbox{cycle}} Idt 
= -\oint (G^1 dX_1 + G^2 dX_2) 
=  \oint \bm{B} \cdot \vec{ds} 
=  \iint  \sigma(X_1,X_2) dX_1 dX_2 
\ee     
In the second expression we define the normal 
vector $\vec{ds}=(dX_2,-dX_1)$ and use the 
notations ${\bm{B}_1=-G^2}$ and ${\bm{B}_2=G^1}$. 
See Fig.4a for an illustration. 
The third expression is obtained via the 
two dimensional version of the divergence 
theorem. If we regard  $\bm{B}$ as a fictitious magnetic 
field, then $\sigma$ is the two dimensional density 
of magnetic charge.

It turns out that in the strict adiabatic 
limit the vector field $\bm{B}$ is 
related to the theory of Berry phase \cite{berry,BR}. 
The formulation of this relation is as 
follows. Assume that the system 
is adiabatically cycled in the ${(X_1, X_2, X_3)}$ space. 
In such case the Berry phase can be 
calculated as a line integral over a 
``vector potential" (also called ``1~form") $\bm{A}$. 
This can be converted by the Stokes theorem into 
a surface integral over a ``magnetic field" 
(also called ``2~form") $\bm{B}$. 
The  $\bm{B}$ field is defined as the ``rotor" of $\bm{A}$. 
It is a divergence-less field but it can have 
singularities  which are known as ``Dirac monopoles".
These monopoles are located at $\vec{X}$ points 
where an occupied energy level has a degeneracy 
with a nearby level.  
Because of ${\Phi\mapsto\Phi+(2\pi\hbar/e)}$ gauge invariance    
the Dirac monopoles form vertical chains 
as illustrated in Fig.4b. Hence we have a distribution  
of what we call ``Dirac chains" \cite{pmc,pme}, 
which is characterized by a density $\sigma(X_1, X_2)$.

\newpage
\subsection{Background and objectives} 
  
Most of the literature about quantum pumping 
deals with the open geometry of Fig.2b. 
The most popular approach is the $S$-scattering  
formalism which leads 
to the B{\"u}ttiker, Pr\^{e}tre and Thomas (BPT) formula \cite{BPT,brouwer}
for the generalized conductance $G$. 
The BPT formula, is essentially a generalization 
of the Landauer formula. In previous publications \cite{pmo,pme}  
we have demonstrated that the BPT formula can be 
regarded as a  special limit of the Kubo formula. 
Our Kubo formula approach to pumping \cite{pmc,pmp} 
leads to ``level by level" understanding of the 
pumping process, and allows to incorporate 
easily non-adiabatic and environmental effects.
In the strict adiabatic limit   
it reduces in a transparent way 
to the theory of adiabatic transport \cite{thouless,avronRev},  
also known as ``geometric magnetism \cite{BR}. 
On the other hand, in the non-adiabatic(!) 
``DC limit" of an open geometry it reduces 
to the $S$ matrix picture, hence resolving 
some puzzles that had emerged in older publications.

The question ``how much charge is pushed 
by translating a scatterer" has been addressed 
in Ref.\cite{avron} in the case of an open 
geometry using the BPT formula.  We have 
addressed the corresponding problem 
of quantum stirring in closed geometry 
in a previous short publication \cite{pmt}, 
but the connection with the Dirac chains  
picture has not been illuminated. 
Furthermore, in \cite{pmt} only the quantum 
chaos limit was considered.

In the present publication we put an emphasis on clarifying 
the route towards quantum-classical correspondence (QCC). 
We shall see that quantum mechanical 
effects are pronounced in {\em simple} systems. 
As the system becomes more chaotic QCC emerges. 
The Dirac chains picture leads to new  
insights regarding the route towards QCC.
These insights are easily missed if we stick 
to the formal Green function calculation of 
our earlier work \cite{pmt}. 
From the above it should be clear that the main objectives 
of the present study are:
\begin{itemize}
\item Derivation of a classical formula for $Q$ (assuming a stochastic picture).
\item Derivation of a quantum result for $Q$ using the Dirac chains picture.
\item Exposing some counter-intuitive results for $Q$ in the case of the simplest models.
\item Illuminating the route towards QCC as we go from ``simple" to ``chaotic" systems.
\end{itemize}
We note that in \cite{pmt} we have presented  
the classical formula for $Q$ without the derivation.

\subsection{Physical motivation and experimental feasibility} 

In the previous section we have explained 
the theoretical motivations for dealing with 
the stirring problem. In the present section 
we would like to further discuss the practicality  
of this line of study, and the feasibility 
of actual experiments. 

It is quite clear that the main focus of today's 
experiments is on {\em open} devices (with leads), 
whereas our interest is in {\em closed} devices. 
Our believe is that {\em ``wireless"} mesoscopic 
or molecular size devices are going to be  
important building blocks of future ``quantum electronics". 
This is of course a vision that people may doubt.    
However, on the scientific side our task is  
to analyze its feasibility.

It is possible to fabricate closed mesoscopic rings, 
and to measure the persistent or the induced currents. 
Experiments with closed devices have been performed 
already~10 years ago. As an example we mention Ref.\cite{orsay} 
where a large array of rings has been fabricated. 
The current measurement has been achieved by coupling 
the rings to a highly sensitive electromagnetic 
superconducting micro-resonator.

The conceptually simplest way to drive a current 
is by inducing an electro motive force (EMF). 
In the setup of Ref.\cite{orsay} the EMF 
has been induced by a ``wire" that spirals on top 
of the array. In our view an attractive alternative option 
would be to induce currents by changing gate voltages 
so as to induce stirring. The advantage of such 
a possibility for the purpose of integrating 
wireless devices in future quantum electronics 
is quite obvious: It is much easier to control 
gate voltages than fluxes of magnetic field.

As far as {\em electronic devices}  
are concerned there is no question about the 
feasibility of realizing quantum stirring by 
manipulating gate voltages, and measuring 
the electrical currents.
But we would like to argue that such possibility 
is open also in case of {\em neutral atoms}.  
It is well known that ``billiards" that confine 
cold atoms can be realized and manipulated \cite{raiz,nir}.
Furthermore, there is no question regarding the 
possibility of creating a ``moving" optical  
barrier so as to create a stirring effect. 
There are variety of techniques to measure the 
induced neutral currents. 
For example one can exploit the Doppler effect 
at the perpendicular direction, which is known as  
the rotational frequency shift \cite{doppler}.

There is one more issue which might be of relevance 
in case of an actual experiment. The Kubo formalism 
assumes that the system settles into a steady state, 
whereas the preparation in case of an actual experiment 
is not very well controlled. We would like to 
argue that the results of the linear response analysis  
are quite robust. This issue is discussed 
in section~4 of Ref.\cite{pms}:  What we get for~$Q$ 
in the Kubo analysis is not merely a formal result, 
but rather a prediction that has an actual physical significance.

\subsection{Outline}

In the first part of this paper  
we review the result for $G$ 
in the case of an open system using the BPT formula.  
Then we present two equivalent derivations 
of the corresponding {\em classical} result 
in the case of a closed geometry. 
We use the term ``classical" in the Boltzmann 
sense. This means that interference within the 
ring is neglected, while the reflection by 
the scatterers (``cross section") is calculated  
quantum mechanically. 
The first derivation is based on a direct solution 
of a master equation, while the second 
is a straightforward application of the Kubo formula.
The classical calculation implies an expression 
for the density $\sigma(X_1,X_2)$ of the monopoles. 
The BPT formula implies $\sigma(X_1,X_2)$ that 
can be regarded as a special case of this calculation.

In the second part of this paper  
we turn to the quantum mechanical analysis. 
As a preliminary stage we discuss the general 
conditions for having a degeneracy point $\vec{X}$ 
in the case of a one dimensional ring.   
Then we review how the pumped charge $Q$ 
can be estimated by calculating a line 
integral that encircles ``Dirac chains". 
Thus we realize that we have to figure out 
what` the distribution $\sigma(X_1,X_2)$ of these 
chains looks like. 
Specifically, we consider the model systems 
that are illustrated in Fig.1, 
and schematically in Fig.2.  
The simplest is a ring where both $g_1$ 
and $g_0$  are modeled as delta barriers. 
The result for $Q$ is quite remote from 
the classical expectation. 
Consequently we try to figure out what happens 
to  $\sigma(X_1,X_2)$  as the system becomes 
more complex: First we add a second fixed barrier, 
and finally we consider what happens 
in the case of a ``chaotic" barrier which 
is modeled using random matrix theory.
We make it clear that the route to 
the classical limit is intimately related 
to so called ``quantum chaos" considerations.

\newpage
\section{Pushing particles in an open geometry}

Let us consider the model of Fig.2b, where 
we have a scatterer within a single mode wire 
which is connected to two reservoirs with 
the same chemical potential.
In this section we assume non-interacting 
spinless electrons and zero temperature Fermi 
occupation.  The scatterer is described by   
\be{0}
V(r;X_1,X_2) = X_2\delta(r-X_1)
\ee
Hence, for some fixed values of $X_1$ and $X_2$ 
its transmission is 
\be{0}
g_0(X_2)=
\left[ 1 
+ \left( \frac{\mathsf{m}}{\hbar^2k_{\tbox{F}}}  X_2 \right)^2 
\right]^{-1}
\ee
where $\mathsf{m}$ is the mass of the particle 
and $k_{\tbox{F}}$ is the Fermi momentum. 
From now on we work with units such that $\hbar=1$. 
The $S$ matrix of the scattering region 
can be written in the general form 
\be{0}
\mathbf{S} = 
\eexp{i\gamma}
\left(\matrix{i\sqrt{1-g}\eexp{i\alpha} & \sqrt{g}\eexp{-i\phi} \cr
\sqrt{g}\eexp{i\phi} &  i\sqrt{1-g}\eexp{-i\alpha}}\right)
\ee
where $\gamma$ is the total phase shift, 
$\alpha$ is the reflection phase shift, 
and $\phi=e\Phi/\hbar$ represents 
the flux which we assume to be zero.   
In the setup of Fig.~2b the length 
of the right lead is $L_A-X_1$ 
and the length of the left lead 
is  $L_B+X_1$. Hence 
\be{0}
g &=& g_0
\\
\gamma &=& k_{\tbox{F}} (L_A+L_B) 
-\arctan\left( \frac{\mathsf{m}}{\hbar^2k_{\tbox{F}}}  X_2 \right)  
\\
\alpha &=& k_{\tbox{F}} (L_A-L_B) 
- 2k_{\tbox{F}} X_1 
\ee
Now that we know the dependence of the $\mathbf{S}$ matrix 
on the parameters $(X_1,X_2)$, the calculation of $G$ 
is quite straightforward. We use the BPT formula    
\be{0}
G^{j} = \frac{e}{2\pi i}
\trc\left(\bm{P}_{\tbox{lead}}\frac{\partial \bm{S}}{\partial X_j}
\bm{S}^{\dag}\right)
\ee
where $\bm{P}_{\tbox{lead}}$ projects on the channels of 
the lead where the current is measured. As indicated in Fig.2b  
the current is measured via a section which is located on 
the right lead. Using the BPT formula we get
\be{0}
G^{1} &=& -(1-g_0) \ \frac{e}{\pi}k_{\tbox{F}}\\
G^{2} &=& - g_0  \ \frac{e}{4\pi\hbar v_{\tbox{F}}} 
\ee
where $v_{\tbox{F}}$ is the Fermi velocity 
corresponding to $k_{\tbox{F}}$. The result 
for $G^1$ is our main interest. It has been 
discussed in Ref.\cite{avron}, 
where the term ``snow plow" has been coined 
in order to describe its physical interpretation. 
Namely, for zero temperature Fermi occupation  
the density of electrons in the wire is ${k_{\tbox{F}}/\pi}$. 
Therefore the number of electrons that are pushed 
by the scatterer is ${dN=(k_{\tbox{F}}/\pi) \times dX_1}$. 
If the transmission of the scatterer is not zero, 
some of the electrons pass through it and consequently 
we have to multiply $dN$ by the reflection probability ${1-g_0}$.

\section{Stirring of particles in a closed geometry}

Let us consider the model of Fig.2a, where the system is closed. 
We assume that the transmission of the ring without 
the moving scatterer is $g^{cl}_1$, while the transmission 
of the scatterer itself is $g_0$. In the following two subsections 
we shall present two optional derivations of the ``classical" 
result for $G$. We use the term ``classical" in the Boltzmann 
sense. Namely, we regard the scattering from 
either $g^{cl}_1$ or $g_0$ as a stochastic process. 
Thus interference within the arms of the ring is 
not taken into account. 
For sake of comparison with the BPT-based result 
we still assume zero temperature Fermi occupation 
(while in later sections we shall allow any arbitrary occupation).   
Within this framework we obtain:
\be{0}
G^{1} &=& 
- \left[\frac{ (1-g_0) g^{cl}_1 }{ g_0 + g^{cl}_1 - 2g_0 g^{cl}_1 } \right]
\ \frac{e}{\pi}k_{\tbox{F}}\\
G^{2} &=& 
- \left[\frac{ (1-g^{cl}_1) g_0 }{ g_0 + g^{cl}_1 - 2g_0 g^{cl}_1 } \right]
\ \frac{e}{4\pi\hbar v_{\tbox{F}}} 
\ee
We note that the amount of charge which is pushed 
by translating a scatterer a distance $\Delta X_1$   
can also be written as \cite{pmt}
\be{13}
Q = -G^{1} \Delta X_1 = 
\left[ \frac{1-g_0}{g_0}\right]
\left[ \frac{g_T}{1-g_T}\right]
\ \frac{e}{\pi}k_{\tbox{F}} \times \Delta X_1
\ee
where $g_T$ is the overall transmission of 
the ring (including the moving scatterer) 
if it were opened: 
\be{0}
\left[ \frac{1-g_T}{g_T}\right] = 
\left[ \frac{1-g_0}{g_0}\right] + 
\left[ \frac{1-g^{cl}_1}{g^{cl}_1}\right]
\ee
As expected the charge $Q$ which is 
transported as a result of an $X_1$ displacement 
depends in a monotonic way on 
the reflection coefficient ${1-g_0}$. 
It monotonically increases from zero, 
and attains half of its maximal value 
for $g_0=g^{cl}_1$.  
A plot of $Q$ versus the ``size" of the 
scatterer is presented in Fig.~5 for three 
representative values of $g^{cl}_1$.   
We also plot $Q$ against $X_2$, 
assuming that the scatterer is modeled as 
a delta function.      

\vspace{10mm}

\mpg{
\Cn{
\putgraph[width=0.4\hsize]{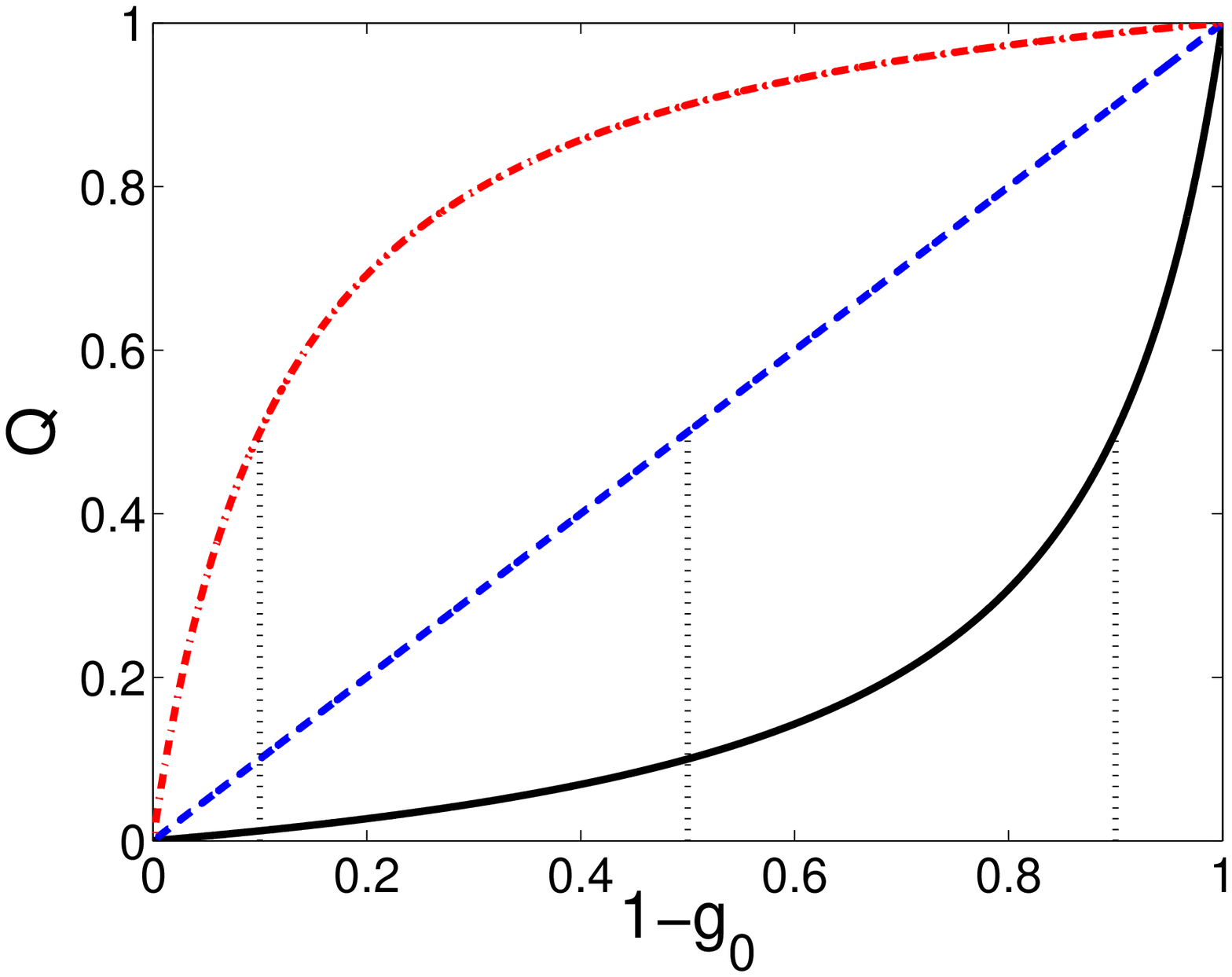}
\putgraph[width=0.4\hsize]{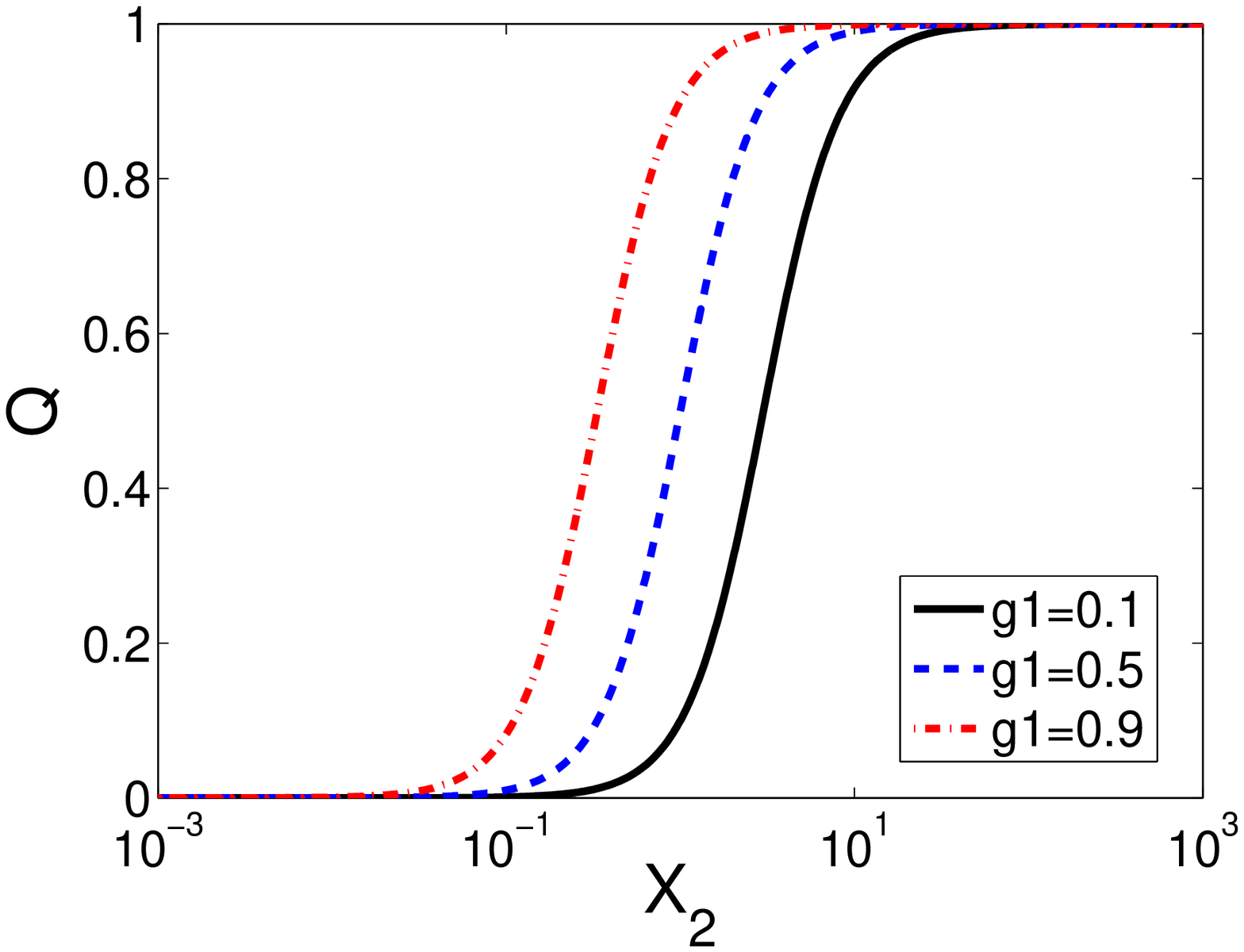}
}

{\footnotesize 
{\bf Fig.5.}
Plots of $Q$ as a function of the ``size" 
of the scatterer. We use arbitrary units 
such that $Q=1$ in the maximum.  
(a) Left panel:  $Q$ is plotted against the reflection 
coefficient ${(1-g_0)}$ for ${g^{cl}_1=0.1}$, 
for ${g^{cl}_1=0.5}$, and for ${g^{cl}_1=0.9}$. 
The dotted lines highlight that $Q$ 
for $g_0=g^{cl}_1$ is half its maximum value.
Note that the BPT based result corresponds 
to ${g^{cl}_1=0.5}$. 
(b) Right panel: Here $Q$ is plotted against $X_2$ 
assuming that the scatterer is a delta function, 
and setting~$\mathsf{m} / (\hbar^2 k_{\tbox{F}}) = 1$.   
}
}

\vspace{3mm}

It is important to realize that the result for an {\em open geometry} 
is formally a special case corresponding to $g^{cl}_1=1/2$. 
This value of $g^{cl}_1$ means that memory is completely 
lost once a particle is scattered by the ``surroundings". 
Namely, if $g^{cl}_1=1/2$ then after a collision a particle 
has equal probability to go in either direction, and any information  
about its initial direction is lost. This observation 
generalizes our discussion in Ref.\cite{kbf} regarding  
the relation between the Kubo and the Landauer conductance.

The classical expression for $G$ implies  
the following result for the density $\sigma(X_1,X_2)$,   
which is illustrated in Fig.6. 
\be{15}
\sigma(X_1,X_2) 
= \frac{d \bm{B}_2 }{dX_2}
= -\frac{e\mathsf{m}}{\pi\hbar^2} 
\frac
{ 2 (1-g^{cl}_1) g^{cl}_1}
{ \left[ 1 
+ \left( 
\left( \frac{\mathsf{m}}{\hbar^2k_{\tbox{F}}} X_2 \right)^2
-1 \right) 
g^{cl}_1 \right]^2 }
\left( \frac{\mathsf{m}}{\hbar^2k_{\tbox{F}}}  X_2 \right)
\ee

\vspace{3mm}

\mpg{
\Cn{
\putgraph[width=0.5\hsize]{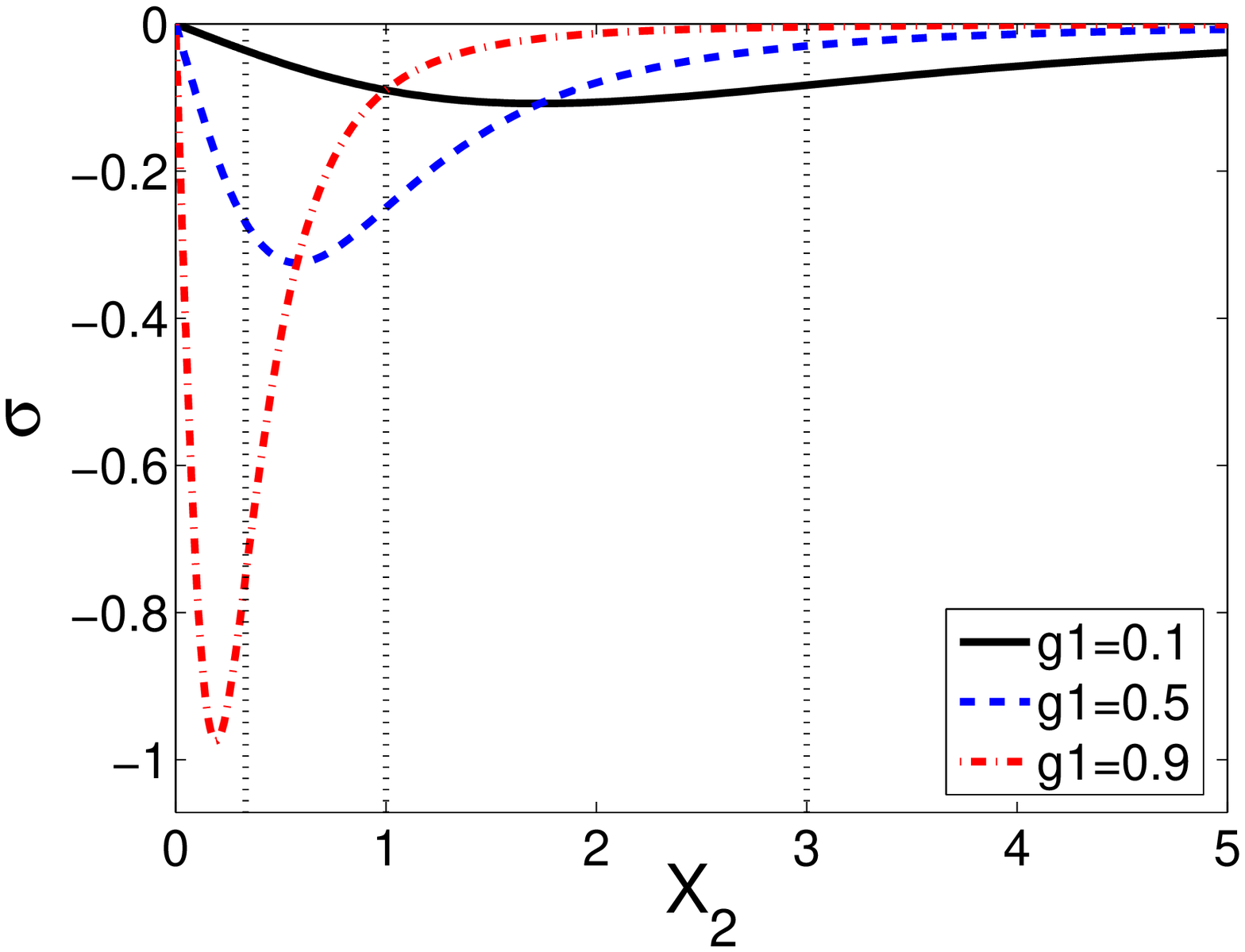}
}

{\footnotesize 
{\bf Fig.6.}
The classically deduced density $\sigma$ 
as a function of ${X_2}$ 
for ${g^{cl}_1=0.1}$, 
for ${g^{cl}_1=0.5}$, 
and for ${g^{cl}_1=0.9}$. 
We use arbitrary units for $\sigma$, 
and set~$\mathsf{m} / (\hbar^2 k_{\tbox{F}}) = 1$.
The dotted vertical lines correspond  
to the median $X_2$ values  
which are determined by the 
equation ${g_0(X_2) = g^{cl}_1}$. 
}
}

\vspace{3mm}

In the following sections we give 
two optional derivations of the classical result. 
The first derivation is based on a physically 
appealing master equation approach, 
in the spirit of the Boltzmann equation.   
The second derivation is a straightforward  
application of the Kubo formula. 
The calculation is done for $G^1$ and can be 
easily modified in order to get $G^2$. 
The advantage of the Kubo formula approach 
is that it can be generalized 
to the quantum mechanical case, 
and it allows the incorporation 
of non-adiabatic and environmental effects.

\section{Classical derivation using a master equation}

We consider a ring with two scatterers: a moving 
scatterer $g_0$ whose velocity is $\dot{X}$,  
and a fixed scatterer $g_1$. A collision of a particle 
with the moving scatterer implies that its velocity 
is changed $v \mapsto v \pm 2\dot{X}$, 
where the sign depends on whether the collision 
is from the right or from the left. 
The associated change in the kinetic energy 
is ${E\mapsto E \pm 2\mathsf{m}v\dot{X} + \mathcal{O}(\dot{X}^2) }$ 
respectively. There are two regions ($x<0$ and $x>0$) 
on the two sides of the $g_0$ scatterer. 
Accordingly we have four distribution functions  
that satisfy the following balance equations: 
\be{0}
\frac{\partial\rho_{+}^{\rightarrow}}{\partial t} 
&=& 
- \left[\rho_{+}^{\rightarrow}v\right]  
+ g_0 \left[\rho_{-}^{\rightarrow}v\right]
+ (1-g_0) \left[\rho_{+}^{\leftarrow}v\right]_{E-2\mathsf{m}v\dot{X}}  
\\  
\frac{\partial\rho_{+}^{\leftarrow}}{\partial t} 
&=& 
- \left[\rho_{+}^{\leftarrow}v\right]  
+ g_1 \left[\rho_{-}^{\leftarrow}v\right]
+ (1-g_1) \left[\rho_{+}^{\rightarrow}v\right] 
\\
\frac{\partial\rho_{-}^{\rightarrow}}{\partial t} 
&=& 
- \left[\rho_{-}^{\rightarrow}v\right]  
+ g_0 \left[\rho_{+}^{\rightarrow}v\right]
+ (1-g_0) \left[\rho_{-}^{\leftarrow}v\right] 
\\
\frac{\partial\rho_{-}^{\leftarrow}}{\partial t} 
&=& 
- \left[\rho_{-}^{\leftarrow}v\right]  
+ g_1 \left[\rho_{+}^{\leftarrow}v\right]
+ (1-g_1) \left[\rho_{-}^{\rightarrow}v\right]_{E+2\mathsf{m}v\dot{X}}  
\ee
The zero order solution in $\dot{X}$ is to have all 
the four distribution functions equal 
to some arbitrary function $f(E)$.  
In the presence of driving, assuming that the 
system has reached a steady state, 
we still have to satisfy the two $\dot{X}$-free 
equations, leading to 
\be{0}
\rho_{+}^{\leftarrow} &=& 
g_1 \rho_{-}^{\leftarrow} + 
(1-g_1) \rho_{+}^{\rightarrow}
\\
\rho_{-}^{\rightarrow} &=& 
g_1 \rho_{+}^{\rightarrow} + 
(1-g_1) \rho_{-}^{\leftarrow}
\ee 
Substitution into the two other equations 
leads after linearization to 
\be{0}
\rho_{+}^{\rightarrow}(E)-\rho_{-}^{\leftarrow}(E) = 
-2\mathsf{m}v\dot{X} \left(\frac{1-g_0}{g_0+g_1-2g_0g_1}\right) 
\frac{\partial f(E)}{\partial E}  
\ee
and for the current we get
\be{0}
I &=& \int_0^{\infty} \frac{dp}{2\pi} 
\ (\rho_{\pm}^{\rightarrow}-\rho_{\pm}^{\leftarrow}) ev
=   \int_0^{\infty} \frac{dp}{2\pi} 
g_1 (\rho_{+}^{\rightarrow}-\rho_{-}^{\leftarrow}) ev
\\
&=&  -\dot{X} 
\int_0^{\infty} 
\left[\frac{e}{\pi}\left(\frac{(1-g_0)g_1}{g_0+g_1-2g_0g_1}\right) \mathsf{m}v \right] 
\frac{\partial f(E)}{\partial E} dE 
\ee 
With the assumption of zero temperature Fermi occupation 
this gives the cited result for $G^1$.

\section{Classical derivation using the Kubo formula}

The generalized fluctuation-dissipation version of 
the Kubo formula (see Ref.\cite{pme} and further references therein)  
relates the generalized conductance to the the cross   
correlation function of the current ${\mathcal{I}}$ and the 
generalized force ${\mathcal{F}}=-\partial\mathcal{H}/\partial X$.
If $X$ is the displacement $X_1$ of the scatterer then 
\be{25}
\mathcal{F} 
= -\frac{\partial\mathcal{H}}{\partial X_1}
= X_2 \delta'(x-X_1)
\ee
For the sake of comparison with previous results 
we assume zero temperature Fermi occupation.
Then the Kubo formula takes the form 
\be{26}
G = \mathsf{g}(E_{\tbox{F}})  \int_0^{\infty} 
\langle {\cal I}(\tau) {\cal F}(0) \rangle  d\tau
= \frac{L}{\pi \hbar v_{\tbox{F}}}  \langle {\cal Q} {\cal F} \rangle
\ee
where $\mathsf{g}(E)=L/(\pi \hbar v_{\tbox{F}})$ 
is the density of states. This density of states 
is proportional to the total ``volume" of the network 
which is $L$.  In the second expression 
we got rid of the time by introducing the notation 
\be{27}
\mathcal{Q} = \int_0^{\infty} {\cal I}(\tau)  d\tau
\ee 
It should be clear that both the generalized force $\mathcal{F}$
and the transported charge $\mathcal{Q}$ are functions 
in phase space, and that ${\langle...\rangle}$ 
stands for phase space average over position and velocity.  
For $\mathcal{F}$ we already have an explicit expression Eq.(\ref{e25}).  
Now we have to figure out what is $\mathcal{Q}$.

On the ring there are two scatterers, 
and one point $x=x_0$  where the current 
is measured. Hence the ring is divided 
into 3 segments. In addition, there are two possible 
directions of motion (clockwise, anticlockwise). 
Hence the phase space is divided into 6 regions. 
It is obvious that the outcome from Eq.(\ref{e27}) 
depends merely on which region the classical 
trajectory had started its journey in.
In fact we need to consider only the 4 regions 
where the particle starts in the vicinity of the 
moving scatterer, else $\mathcal{F}$ vanishes.
So we have the $``+"$ region between the moving 
scatterer and $x_0$, and the  $``-"$ 
region on the other side between the two scatterers.   
Accordingly the four possible outcomes from Eq.(\ref{e27}) are: 
\be{28}
Q_{+}^{\rightarrow} &=& e\left[\frac{1}{2(1-g_T)}\right]
\\ \label{e29}
Q_{+}^{\leftarrow} &=& - e\left[\frac{1}{2(1-g_T)} - 1 \right]
\\ \label{e30}
Q_{-}^{\rightarrow} &=& \left[\frac{g_0}{1-(1-g_1)(1-g_0)} 
-\frac{g_1(1-g_0)}{1-(1-g_1)(1-g_0)}\right] Q_{+}^{\rightarrow} 
=\frac{g_0-g_1+g_0g_1}{g_0+g_1-g_0g_1} Q_{+}^{\rightarrow}
\\ \label{e31}
Q_{-}^{\leftarrow} &=& -\frac{g_1-g_0+g_0g_1}{g_0+g_1-g_0g_1} Q_{+}^{\rightarrow}
\ee
The derivation of the above expressions is as follows.
It is simplest if the particle starts in the $``+"$ region, 
because then we can regard the two scatterers as one 
effective scatterer $g_T$.  Assume that at time $t=0$ the 
particle approach $x=x_0$ from the left. 
The charge that goes through the section after a round trip 
is suppressed by a factor $(2g_T-1)$ 
due to the scattering (we sum the clockwise 
and the anticlockwise contributions). 
Thus we find that the total charge that 
goes through the section due to multiple 
reflections is a geometric sum that leads 
to Eq.(\ref{e28}). If we start in the $``+"$ 
region in the opposite direction, then we have 
the same sequence but with the opposite sign 
and without the first term. Hence we get Eq.(\ref{e29}).   
Next assume that at $t=0$ the particle starts in the $``-"$ 
region, and approaches $g_0$ from the left. 
Then we can have at a later time 
a positive pulse of current. The probability 
for that is the geometric summation over 
$g_0 ((1-g_0)(1-g_1))^{\tbox{integer}}$. 
Otherwise, we get a negative pulse of 
current, with a complementary probability 
that can be regarded as a geometric summation over 
$g_1 ((1-g_0)(1-g_1))^{\tbox{integer}} (1-g_0)$. 
Thus the total current through the section,  
taking into account all subsequent multiple reflections 
(rounds) is given by Eq.(\ref{e30}). 
A similar calculation leads to Eq.(\ref{e31}).

Since there are only four possible values for $\mathcal{Q}$ 
the calculation of the phase space average becomes trivial: 
\be{-1}
\langle {\cal Q} {\cal F} \rangle 
= \frac{1}{2L} \left[ \int_{+}\mathcal{F}dr \right] Q_{+}^{\rightarrow}
+ \frac{1}{2L} \left[ \int_{+}\mathcal{F}dr \right] Q_{+}^{\leftarrow}
+ \frac{1}{2L} \left[ \int_{-}\mathcal{F}dr \right] Q_{-}^{\rightarrow}
+ \frac{1}{2L} \left[ \int_{-}\mathcal{F}dr \right] Q_{-}^{\leftarrow} 
\ee
The integral over $\mathcal{F}$ is taken either 
within the $``+"$ or within the $``-"$ region. It is 
trivially related to the momentum impact 
and yields the result
\be{0}
\int_{\pm}\mathcal{F}dr = \mp \mathsf{m}v_{\tbox{F}}^2
\ee
Putting everything together we get the desired result for $G^1$. 
With some minor modifications we can calculate $G^2$ using 
the same procedure.

\section{The quantum mechanical picture}

The Kubo formula holds also in the quantum mechanical case. 
But now $\mathcal{I}$ and $\mathcal{F}$ are operators,
so it is more convenient to express the Kubo formula 
using their matrix elements. After some algebra one 
obtains the result: 
\be{33}
G = \sum_{m(\ne n)}
\frac{ 2\hbar  \im[\mathcal{I}_{nm}] \mathcal{F}_{mn} }
{(E_m-E_n)^2 + (\Gamma/2)^2}
\ee
For more details see Ref.\cite{pme} and further references therein.  
In the above formula it is assumed that only one energy level ($n$) 
is occupied. If we have zero temperature Fermi occupation, 
then we have to sum over all the occupied levels. 
The Kubo formula incorporates a parameter~$\Gamma$ that 
reflects either the non-adiabaticity of the driving, 
or environmentally induced ``memory loss" due to decoherence.   
For a strictly isolated system in the strict adiabatic limit  
we have $\Gamma=0$. Then we identify $G$ as an element 
of Berry's field $\bm{B}$,  
as explained in the introduction. The effect of $\Gamma$ 
on $\bm{B}$ will be discussed below.

We would like to see how the classical result can emerge 
in some limit from the above quantum expression. 
It turns out that this does not require a detailed calculation. 
We can use some topological properties of $\bm{B}$ 
in order to figure out the answer!  The main observations 
that we further explain below are: 
\begin{itemize}
\item[\bf{(1)}] The $\bm{B}$ is divergence-less with the exception of Dirac monopoles  
\item[\bf{(2)}] The monopoles are arranged in $\vec{X}$ space as vertical chains  
\item[\bf{(3)}] The far field of $\bm{B}$ is like a two-dimensional electrostatic problem 
\item[\bf{(4)}] Only non-compensated chains give net contribution   
\end{itemize}
As long as the occupied level $n$ does not have a degeneracy with 
a nearby level,  $\bm{B}$ is finite and divergence-less. 
Only at degeneracies can it become singular. It can be argued     
that these singularities must have their charge quantized in units 
of $\hbar/2$ else the Berry phase would be ill defined.
We have defined $X_3=\Phi$ as the Aharonov-Bohm flux through the ring. 
This means that if we change $X_3$ by $2\pi\hbar/e$  then by 
gauge invariance we have another degeneracy.  This means 
that the Dirac monopoles are arranged as vertical chains, 
and that the average charge per unit length is $e/(4\pi)$.
Thus the far field of a Dirac chain is as in a two dimensional 
electrostatic problem. If we calculate the line integral  
of Eq.(\ref{e1}) then we get, within the framework of the far field 
approximation, $Q=1$. Thus we conclude that if we have several 
Dirac chains of the same ``sign", then $Q$ simply counts how 
many are encircled. 

We have to notice that if we have 
Fermi occupation, then the {\em net} contribution comes only 
from degeneracies of the last occupied level with the first 
unoccupied level. This is what we meant above (item 4) 
by ``non-compensated". In order to avoid 
misunderstanding of the ``compensation" issue 
let us discuss with some more details what happens 
if two neighboring levels $n$ and $m$ are occupied.
With the level $n$ we associate a field $\bm{B}^{(n)}$, 
while with $m$ we associate a field $\bm{B}^{(m)}$. 
In general $\bm{B}^{(m)} \ne -\bm{B}^{(n)}$. If we are near 
a degeneracy than we may say that $\bm{B}^{(n)}$ 
emerges from a Dirac chain which is associated 
with level $n$, while $\bm{B}^{(m)}$ emerges from 
a Dirac chain which is associated with level $m$. 
By inspection of Eq.(\ref{e33}), taking into account 
that ${\im[\mathcal{I}_{nm}]=-\im[\mathcal{I}_{mn}]}$, 
we realize that the two Dirac chains have opposite 
charge. Their corresponding fields do not cancel  
each other, but the total field is no longer singular, 
implying that the {\em net} charge is zero.

In the quantum stirring problem we shall see that the 
$X_1$ distance between non-compensated chains is simply 
half the De-Broglie wavelength $\lambda_E=2\pi/k_E$. 
From this it follows that the amount of charge which is pushed 
by a very ``large" scatterer is  
\be{34}
Q \approx 
e \frac{\Delta X_1}{\lambda_{\tbox{E}}/2}
= e \frac{k_{\tbox{E}}}{\pi} \times \Delta X_1 
\ee
What happens if the cycle is not in the ``far field" 
but rather passes through the distribution of the monopoles? 
To be more specific let us consider what happens 
to $Q$ if we displace the scatterer a distance $\Delta X_1$. 
What is the dependence on $X_2$?
Do we get the classical result as in Fig.5?
Obviously, in order to get the classical result 
the distribution $\sigma(X_1,X_2)$ 
should be in accordance with Eq.(\ref{e15}). 
Strictly speaking this is {\em not} the case 
because we have a discrete set of monopoles rather 
than a smooth distribution of ``magnetic charge". 
Still we can hope that $\sigma(X_1,X_2)$
would be classical-like upon course graining. 
We discuss further this issue in the next paragraphs.

If we make a pumping cycle in the vicinity of a monopole 
then it is obvious that the result would be 
very different from the classical prediction. 
What we expect to get in the quantum mechanical case is 
illustrated in the upper panel of Fig.7.
For a cycle that goes very close to a monopole 
the charge can be huge. In reality it is very difficult 
to satisfy the adiabatic condition near a degeneracy, 
or else there are always environmental effects. 
Either way, once we have a finite $\Gamma$, the result 
that we get for $Q$ is smoothed.

If the pumping cycle passes through a distribution 
of many monopoles then what we expect to get  
(as we deform or shift the cycle) are huge 
fluctuations as illustrated in the lower panel of Fig.7.   
Again, the effect of either non-adiabaticity or environmental 
effects is to smooth away these fluctuations. 
The interested reader can find some further discussion 
of this point including a numerical example in \cite{pmt}.

\vspace{3mm}

\mpg{
\Cn{
\putgraph[width=0.3\hsize]{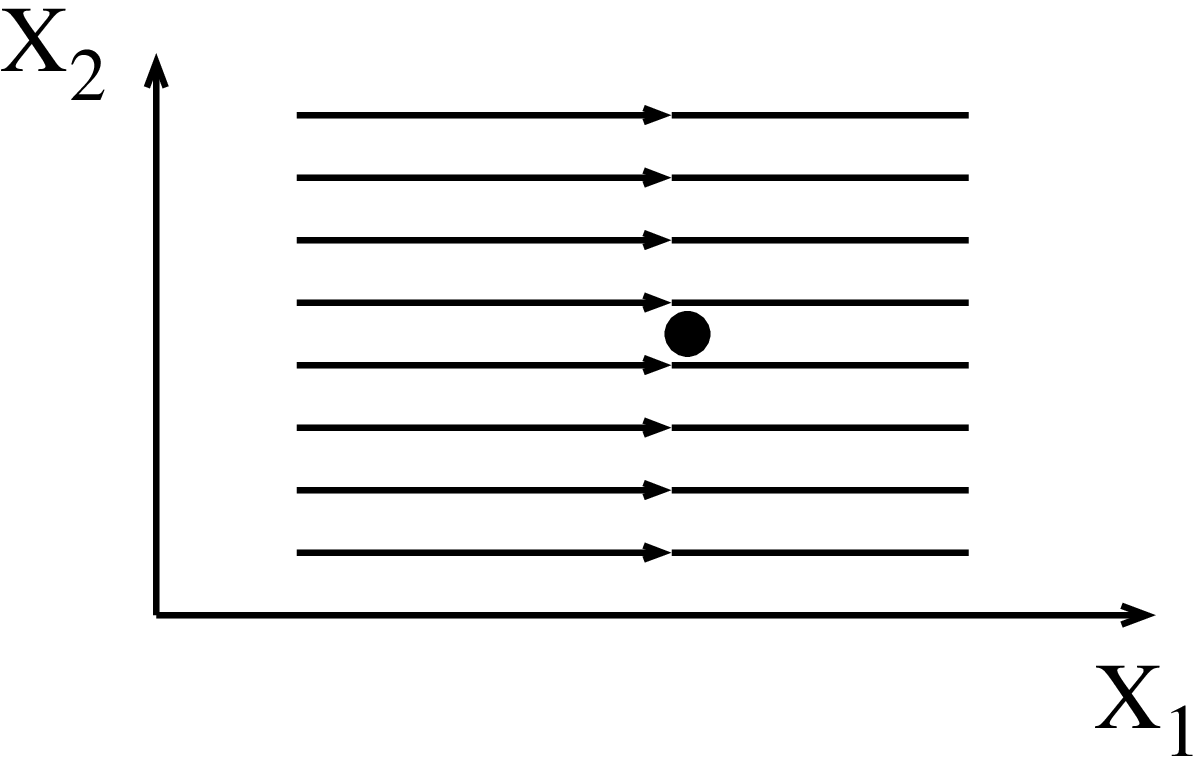}
\hspace{0.05\hsize}
\putgraph[width=0.4\hsize]{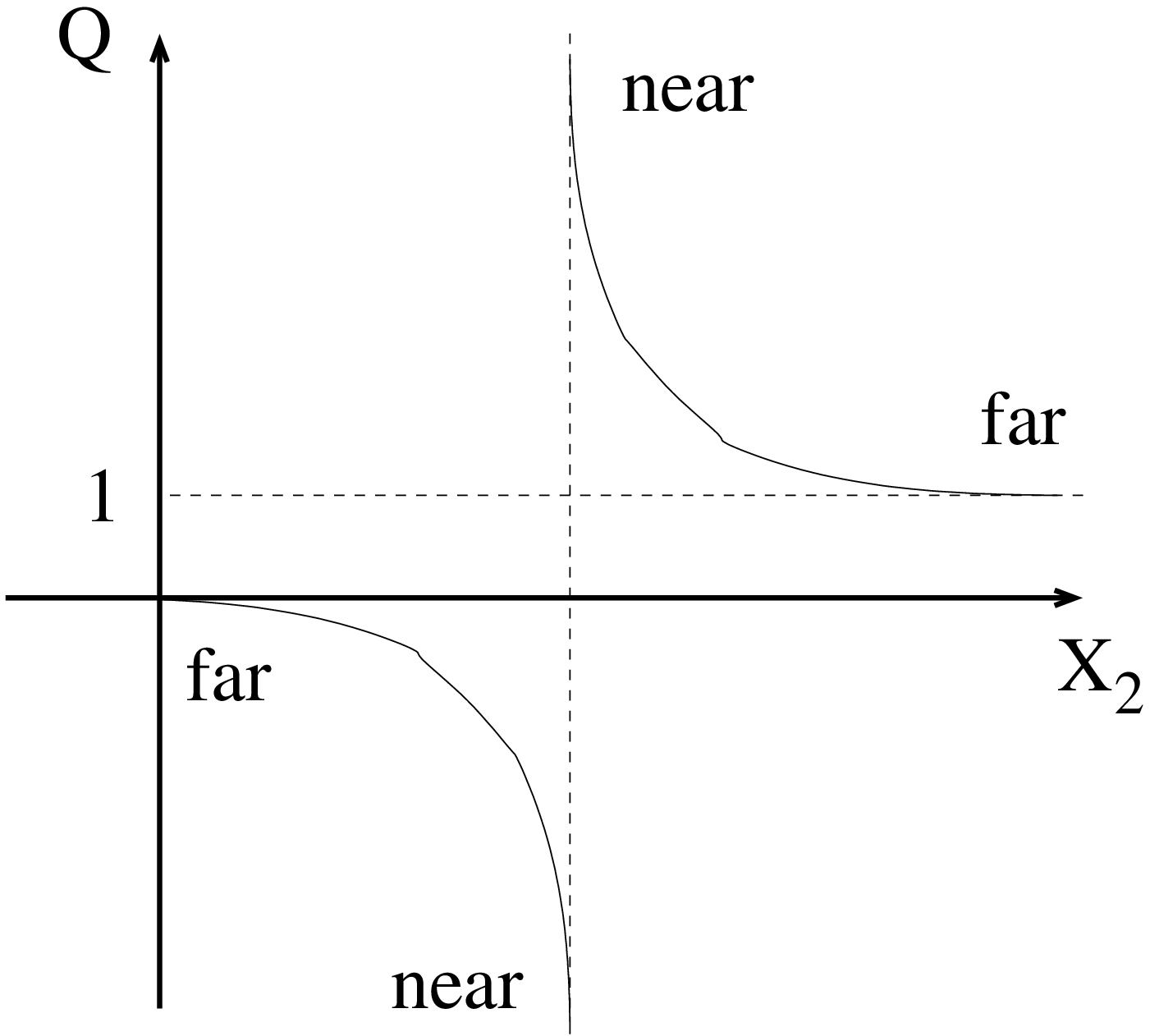} 
\\ \vspace{5mm}
\putgraph[width=0.3\hsize]{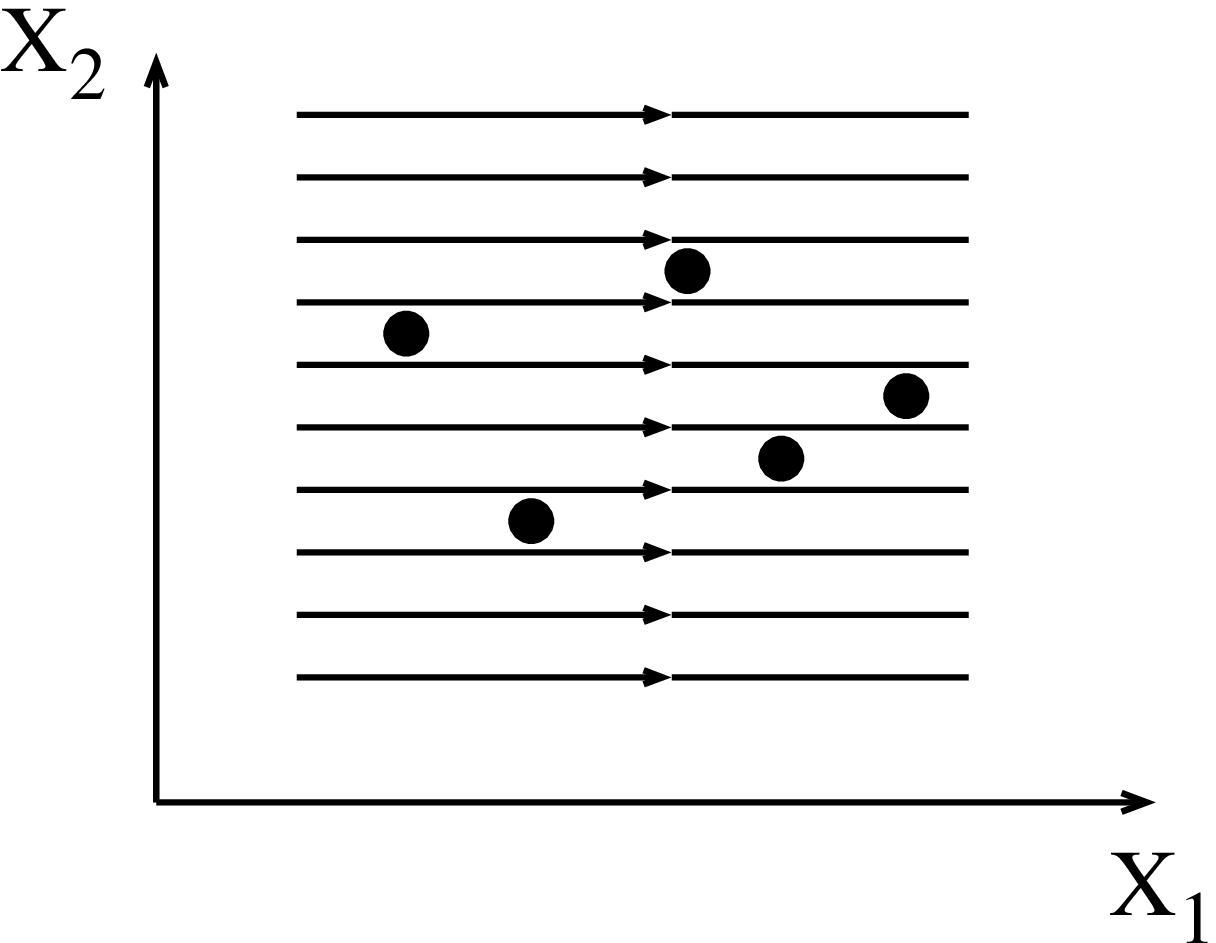} 
\hspace{0.05\hsize}
\putgraph[width=0.4\hsize]{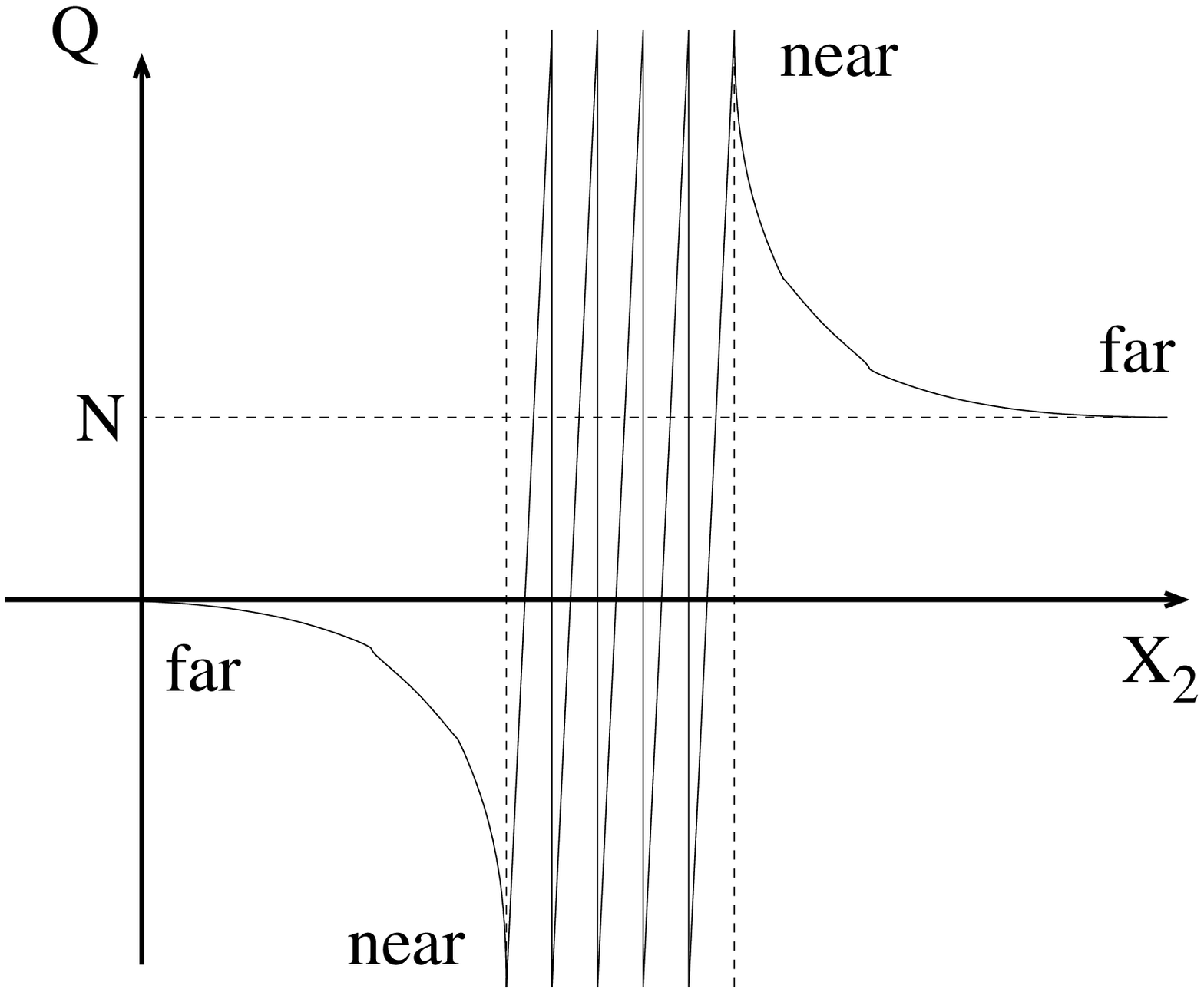} 
}

{\footnotesize 
{\bf Fig.7.}
Several pumping cycles are indicated  
in the left panels: It is implicit that 
each segment is closed as in Fig.3. 
The black points represent degeneracies. 
For each pumping cycle one can calculated $Q$. 
The qualitative expectation for the outcome 
is illustrated in the right panels. 
In the upper illustration we assume 
that the pumping cycle encircles only 
one degeneracy, while in the lower 
illustration we assume that it encircles $N$ 
degeneracies. In a later section we 
display numerical results that support 
the illustrated expectations.
}
}

\vspace{3mm}

Coming back to the quantum-classical correspondence (QCC) 
issue, we realize that at best QCC can be satisfied  
in a statistical sense.  So we ask whether the 
coarse grained $\sigma(X_1,X_2)$ agrees 
with the classical expectation Eq.(\ref{e15}).
The answer which we give in the following sections, 
is that QCC is not realized in the case of simple 
non-chaotic models. In the ``simple" cases we 
get a non-classical $\sigma(X_1,X_2)$ 
and hence a different dependence of $Q$ on $X_2$.

\section{The degeneracies in $X$ space}

We can use the scattering approach in order to find the 
energy levels of a ring. In this approach the ring is opened 
at some arbitrary point and the $S$ matrix of the 
open segment is specified. It is more convenient to  
use the row-swapped matrix, such that the transmission amplitudes 
are along the diagonal:
\be{0}
\tilde{\mathbf{S}}(E;X_1,X_2) = 
\eexp{i\gamma}
\left( \amatrix{\sqrt{g}\eexp{i\phi} &  i\sqrt{1-g}\eexp{-i\alpha} \cr
i\sqrt{1-g}\eexp{i\alpha} & \sqrt{g}\eexp{-i\phi}} \right)
\ee
The periodic boundary conditions imply the following 
secular equation      
\be{0}
\det(\tilde{\mathbf{S}}(E;X_1,X_2) -\bm{1}) = 0
\ee
Using
\be{0}
&& \det(\tilde{\mathbf{S}}-I) = \det(\tilde{\mathbf{S}})-\trc{(\tilde{\mathbf{S}})}+1 \\
&& \det(\tilde{\mathbf{S}}) = (\eexp{i\gamma})^2  \\
&& \trc{(\tilde{\mathbf{S}})} = 2\sqrt{g}\eexp{i\gamma}\cos{\phi}
\ee
we get 
\be{0}
\cos(\gamma(E)) = \sqrt{g(E)} \cos(\phi)
\ee
In order to find the eigen-energies we plot both sides as 
a function of $E$. The left hand side oscillates between $-1$ 
and $+1$,  while the right hand side may have a smaller 
amplitude. It is not difficult to realize that the only way 
to have two eigen-energies coincide is to get  
\be{0}
\begin{array}{cc}
\left\{ \begin{array}{ccc}
\phi &=& 0 \ \mbox{mod}(2\pi)\\
g &=& 1 \\ 
\gamma &=& n_{\tbox{even}}\pi
\end{array}
\right\}
\hspace{0.05\hsize} \mbox{or} \hspace{0.05\hsize}
\left\{ \begin{array}{ccc}
\phi &=& \pi \ \mbox{mod}(2\pi)\\
g &=& 1 \\ 
\gamma &=& n_{\tbox{odd}}\pi
\end{array}
\right\}
\end{array}
\ee
where $n$ is either even or odd integer that can 
be exploited (if we keep track over $\gamma$) as a level counter.

Both $g$ and $\gamma$ depend on $(E;X_1,X_2)$. 
Since we want $g$ to be maximal the condition 
for having a degeneracy involves 4 rather than 3 equations 
as we are going to see below. 
An immediate conclusion is that we have 
two types of Dirac chains: those that have monopoles 
in the plane of the pumping cycle ($X_3=\Phi=0$), 
and the others that have monopoles off 
the plane of the pumping cycle.

In our model system we have two scatterers. 
One is the moving scatterer and the other 
is the rest of the network. The two are connected 
by arms of length $L_A-X_1$ and  $L_B+X_1$. 
The constants $L_A$ and $L_B$ can be absorbed into the definition 
of the surrounding network. Each scatterer is fully characterized 
by the set of parameters ${ \{ g_i, \gamma_i, \alpha_i, \phi_i \} }$.
Note that we do not absorb $X_1$ into the definition 
of $\alpha_0$.  After some algebra we find the following expressions 
for the transmission coefficient and for the total phase shift: 
\be{0}
g &=& 
\frac{g_0 g_1}
{2 - g_0 - g_1  +  g_0 g_1 + 2\sqrt{(1-g_0)(1-g_1)}
\cos(\gamma_0 + \gamma_1 + \alpha_0 + \alpha_1 - 2k_E X_1)}
\\
\gamma &=& \gamma_0+\gamma_1 
\ee
where $k_E$ is the wavenumber that corresponds to the energy $E$.  
Thus the conditions for having a degeneracy take the form   
\be{43}
\left\{ \begin{array}{l}
X_3 = \mbox{integer flux} \\
g_0(X_2) = g_1 \\ 
\alpha_0 + \alpha_1 - 2 k_E X_1 = \pi \ \mbox{mod}(2\pi) \\
\gamma_0 + \gamma_1 =  n_{\tbox{even}}\pi
\end{array}
\right.
\left\{ \begin{array}{l}
X_3 = \mbox{half integer flux} \\
g_0(X_2) = g_1 \\ 
\alpha_0 + \alpha_1 - 2 k_E X_1 = 0 \ \mbox{mod}(2\pi) \\
\gamma_0 + \gamma_1 = n_{\tbox{odd}}\pi
\end{array}
\right.
\ee
We have highlighted the dependence on the parameters $(X_1,X_2,X_3)$. 
There is of course also an implicit dependence 
of ${ \{ g_i, \gamma_i, \alpha_i \} }$ on the energy $E$. 
The conditions that are listed above are very intuitive: 
The system should have time reversal symmetry; 
The barriers should ``balance"  each other;  
The phases which are associated 
with the reflections should lead to destructive interference;  
And the total phase shift should respect the periodic boundary conditions.   
 
From Eq.(\ref{e43}c) we see that in general 
the $X_1$ distance between degeneracies that belong to the same level  
is roughly half the De-Broglie wavelength as stated previously. 
The question that we would like to address is how these degeneracies are 
distributed with respect to $X_2$.

\section{Quantum stirring in simple rings}

We would like to find the distribution 
of degeneracies with respect to $X_2$ 
in the simplest model: a ring with two 
delta scatterers (see Fig.1). The arms 
that connect the two scatterers are 
of length $L_A+X_1$ and $L_B-X_1$. 
For the $S$ matrix that represents 
the fixed scatterer (including the arms) we have 
\be{0}
g_1(E) &=& 
\left[ 1 + \left( \frac{\mathsf{m}}{\hbar^2k_{\tbox{F}}} V \right)^2 \right]^{-1}
\\ 
\gamma_1(E) &=& k_E (L_A+L_B)
-\arctan\left( \frac{\mathsf{m}}{\hbar^2k_{\tbox{F}}}  X_2 \right)
\\ 
\alpha_1(E) &=& k_E (L_A-L_B)
\ee
Since the dependence of $g_0$ and $g_1$ on the 
barrier ``size" has the same functional form,  
the condition Eq.(\ref{e43}c) implies $X_2=V$ 
irrespective of $E$.
Thus we get that all the degeneracies  
are concentrated at the same $X_2$. 
This is clearly very different from 
the classically expected distribution.

In Fig.8 we display an example. The degeneracies 
that are associated with the first 7 levels are 
indicated. Filled circles stand for $\phi=0$ degeneracies, 
while hollow circles stand for $\phi=\pi$ degeneracies. 
Only the last (7th) level contributes non-compensated monopoles. 
The $X_1$ distance between the non-compensated monopoles 
is roughly half De-Broglie wavelength.

In Fig.9 we show what happens to the degeneracies 
if we add a second fixed scatterer. We have chosen 
an additional scatterer that can be treated 
as a perturbation. The calculation was done 
using perturbation theory. 
We shall not present the details 
of this lengthy calculation here. 
For larger perturbations (not presented) we had 
to solve the secular equation numerically. 
This was done using an efficient
algorithm \cite{holger}. In any case, the purpose
of Fig.9 is merely to demonstrate that once the
symmetry of the system is broken the degeneracies
spread out in the $X_2$ direction.

\vspace{3mm}

\mpg{
\Cn{
\putgraph[width=0.49\hsize]{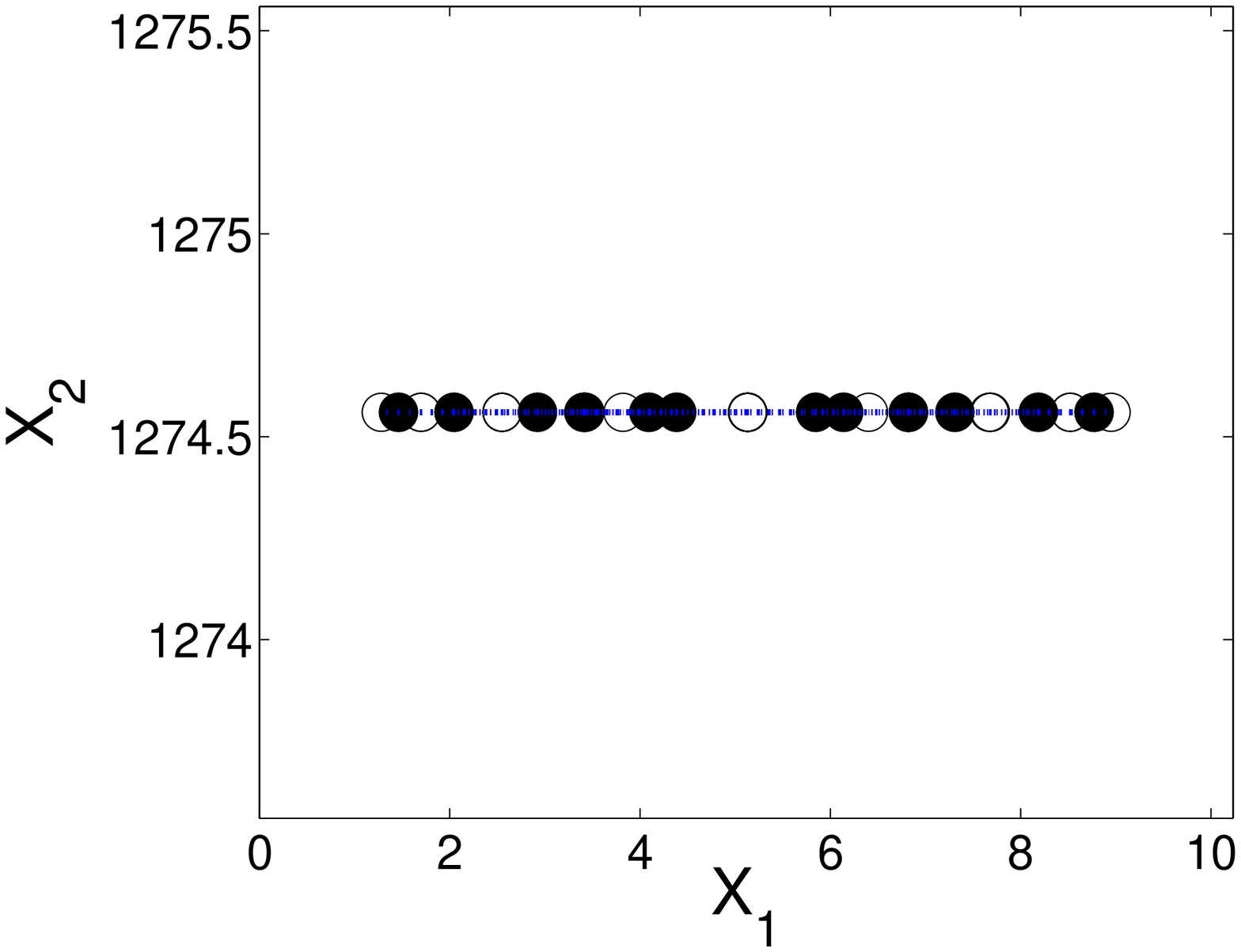}
\putgraph[width=0.49\hsize]{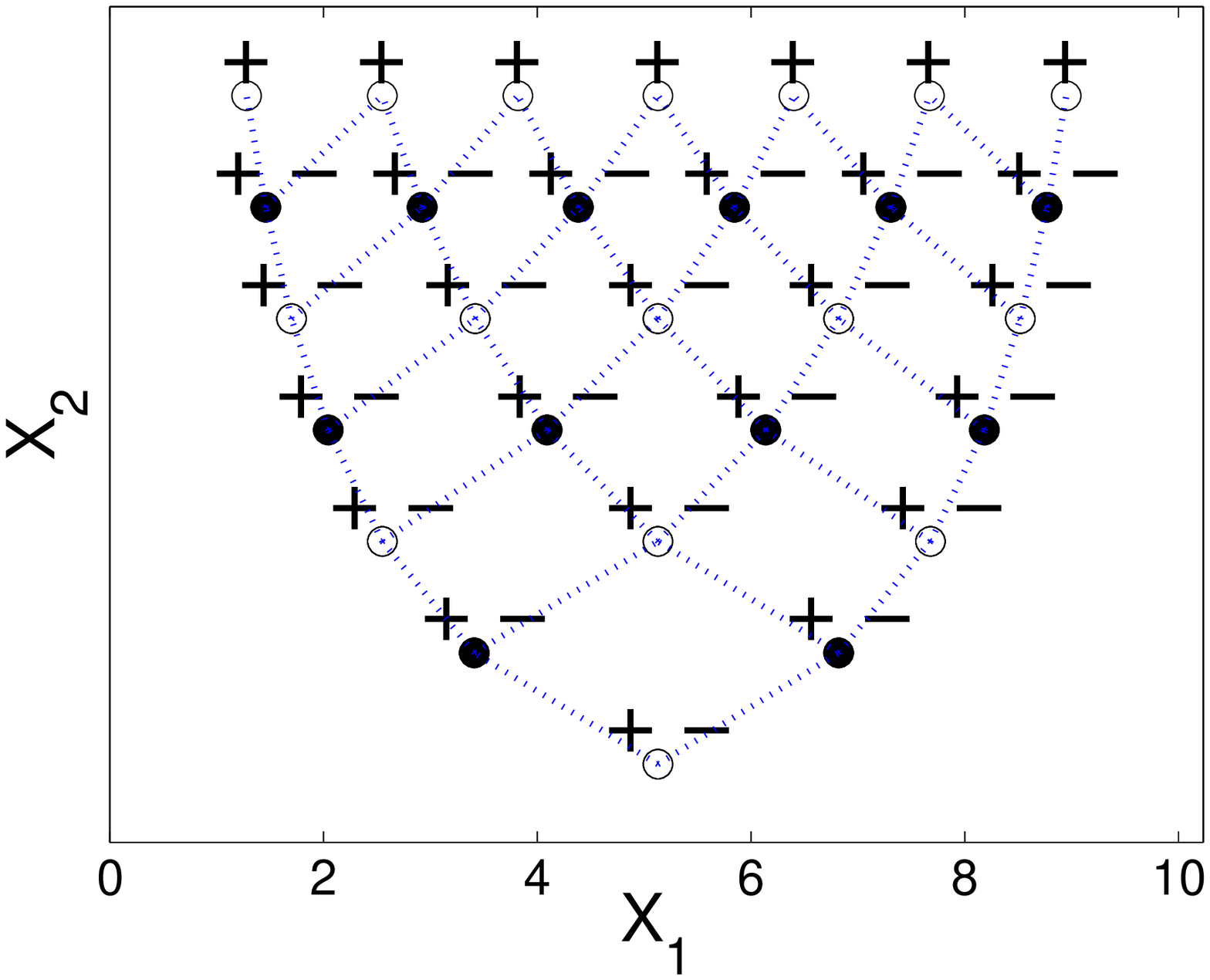}
}

{\footnotesize 
{\bf Fig.8.}
The degeneracies in the double delta model of Fig.1.
We set $L_A=10.23$ and $L_B=0$, so that $X_1$ measures 
the distance from the fixed scatterer. 
The ``size" of the fixed delta scatterer is $V=1274.56$.   
We use units such that $\mathsf{m}=\hbar=1$.
We assume that only the lower~7~levels are occupied.
The filled circles are degeneracies on the flux zero plane 
and the empty circles are degeneracies on the 
flux $\pi$ plane. The left graph shows the actual 
arrangement in the $(X_1,X_2)$ plane. Namely, 
all the degeneracies are on the line ${X_2=V}$. 
In the right graph the degeneracies were displaced 
for the sake of clarity. Only the~7th occupied level 
contributes non-compensated monopoles.  
}
}

\vspace{3mm}

\mpg{
\Cn{
\putgraph[width=0.6\hsize]{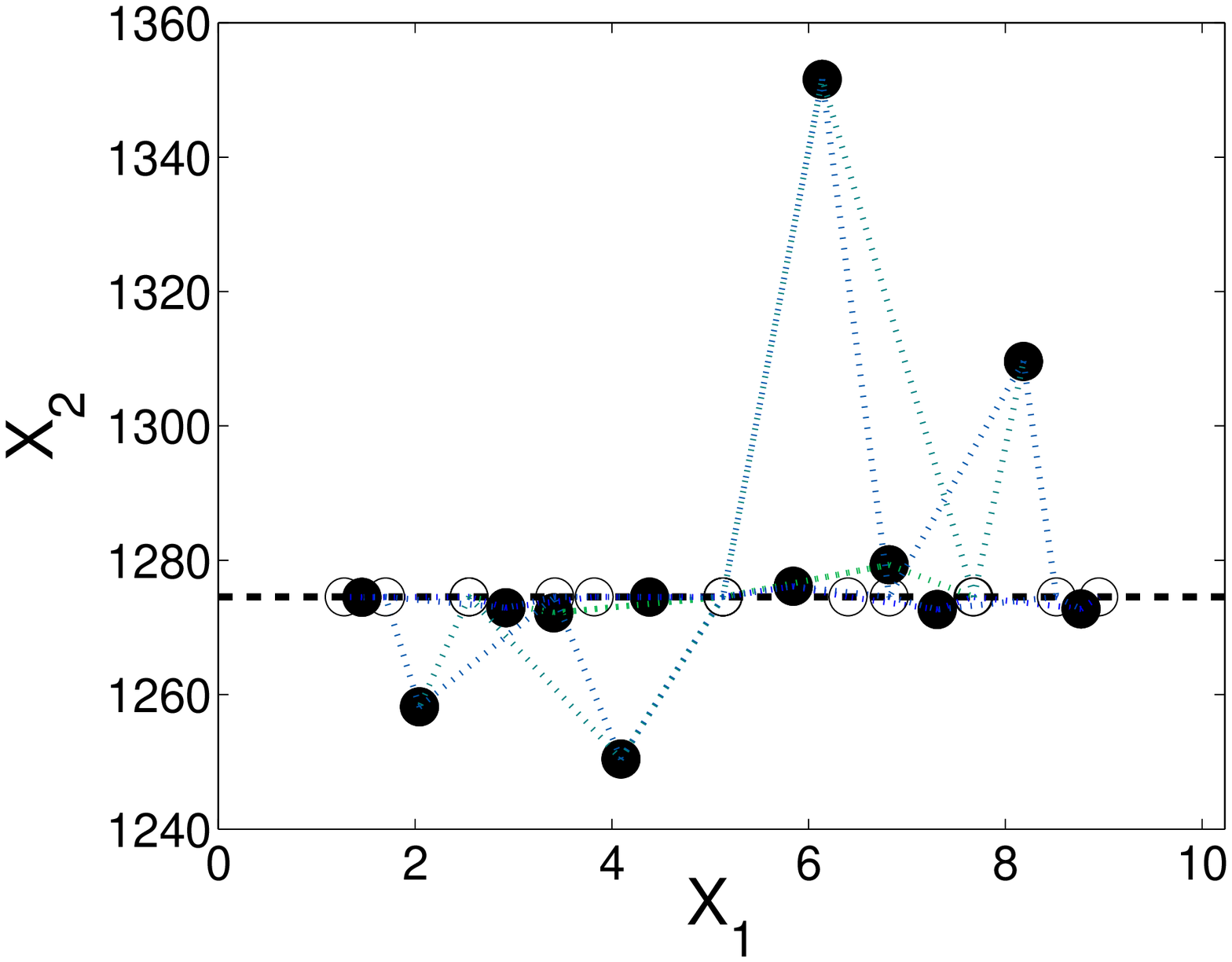}
}

{\footnotesize 
{\bf Fig.9.}
The degeneracies in the triple delta model of Fig.1.
Namely, to the model of Fig.8 we have added a delta
barrier of ``size" $V_P=10^{-5}$, located at ${x=7.61}$.
This additional delta barrier can be treated as a small
perturbation. As a result of this perturbation the degeneracies  
shift and spread out in the $X_2$ direction. 
Degeneracies that belong to the same level 
are connected by a line. As in the previous figure 
only the~7th occupied level contributes non-compensated monopoles. 
}
}

\vspace{3mm}

The distribution $\sigma(X_1,X_2)$ in the case of a ring with 
a single fixed scatterer is very different from the 
classical prediction. Consequently also $Q$ comes
out very different from Eq.(\ref{e15}) [and see also Fig.6].
The reader might be curious to know how $Q$ depends 
on the ``size" ($X_2$) of the scatterer in the case of   
Fermi occupation. So we have calculated $G$ 
numerically using Eq.(\ref{e33}), and integrated 
over it to get $Q$. The numerical results are displayed in Fig.10.     
Further analysis of the crossover 
from ``near field" to ``far field" cycles 
will be published in a separate work \cite{pms}.

\vspace{3mm}

\mpg{
\Cn{
\putgraph[width=0.49\hsize]{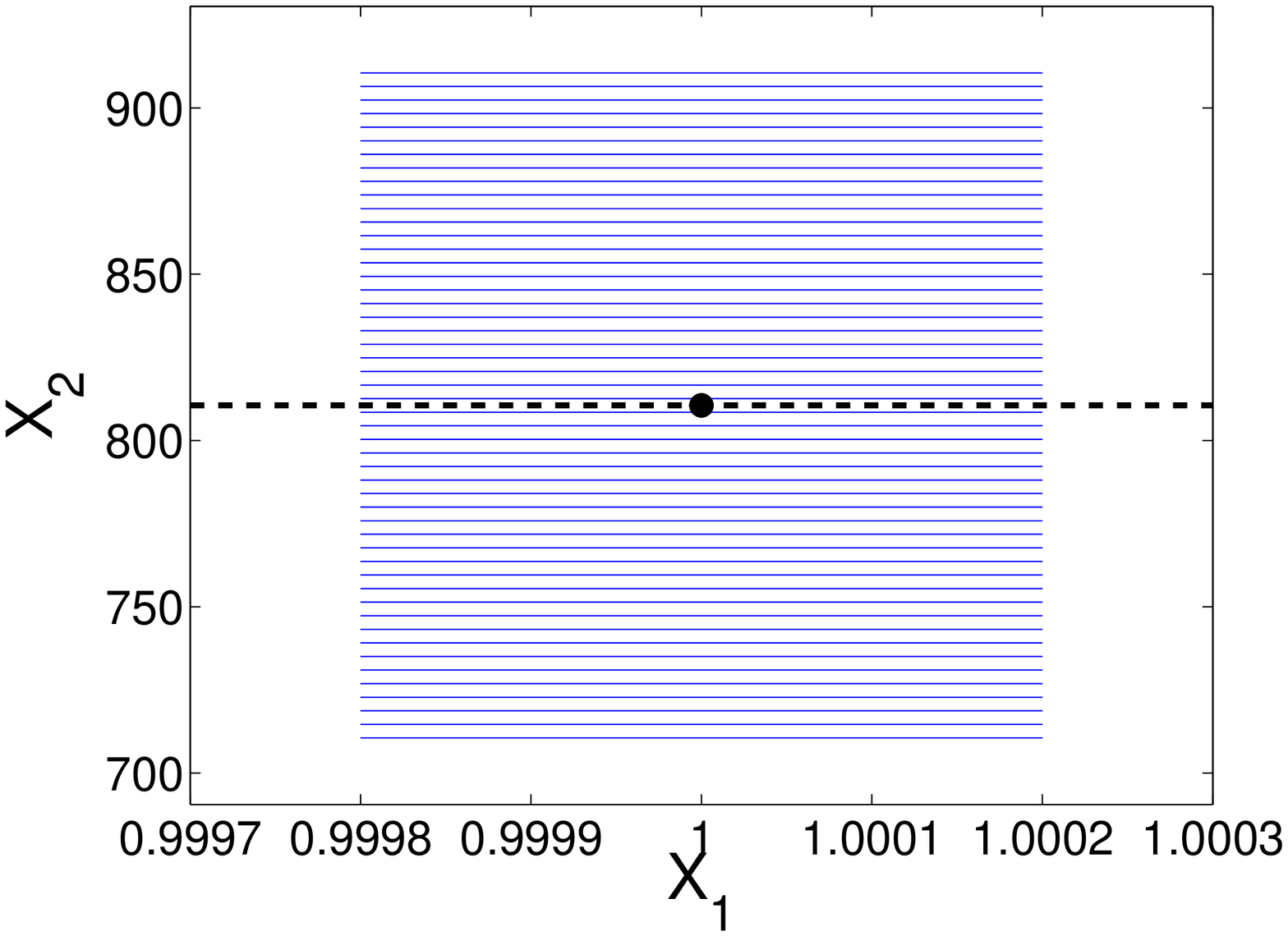}
\putgraph[width=0.49\hsize]{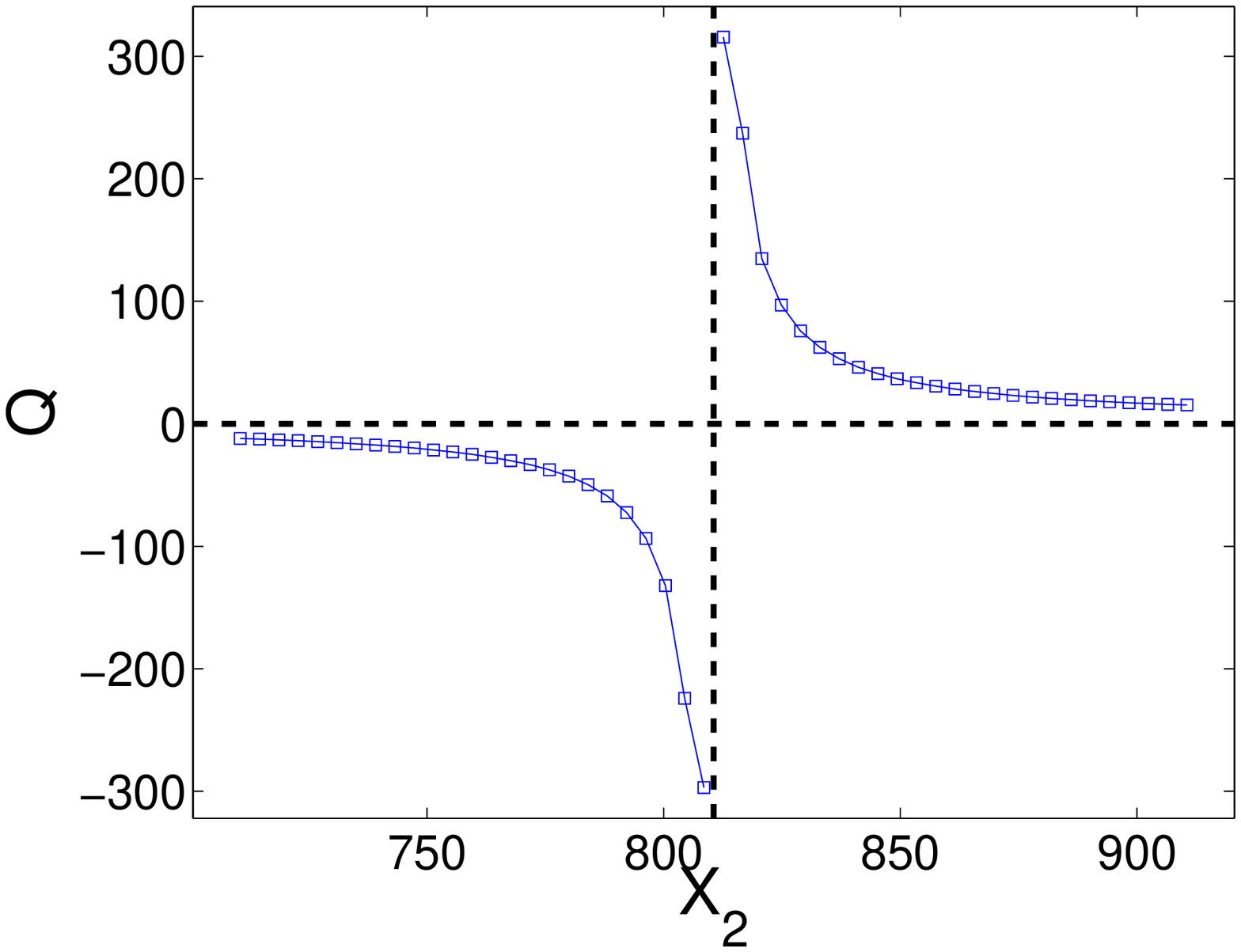}
}

{\footnotesize 
{\bf Fig.10.}
Several pumping routes are displayed 
in the left panel. For each of them $Q$ 
has been calculated numerically.  
The results are displayed in the right panel. 
Note the agreement with the qualitative 
expectation that has been expressed in Fig.7.
The calculation is done for the double 
delta model of Fig.1 with $L_A=1000.23$  
and $L_B=0$. The ``size" of the fixed 
barrier is $V=810.56$. The energy level 
involved are $n=998$ and $m=999$. 
We use units such that $\mathsf{m}=\hbar=1$.
}
}

\section{Quantum stirring in chaotic rings}

We would like to find the distribution 
of degeneracies with respect to $X_2$ 
in case of a chaotic network (see an example in Fig.1).
Let us try to extend the approach that 
has been used in the previous section.
A hypothetical illustration of $g_1(E)$ in the chaotic    
case is displayed in Fig.11. The universal conductance 
fluctuations of $g_1$ are characterized by a one parameter  
probability distribution $P(g_1;\bar{g}_1)$ which 
we discuss below. This probability distribution 
depends on one parameter, which we choose to be 
the average transmission$\bar{g}_1$.

\vspace{3mm}

\mpg{
\Cn{
\putgraph[width=0.4\hsize]{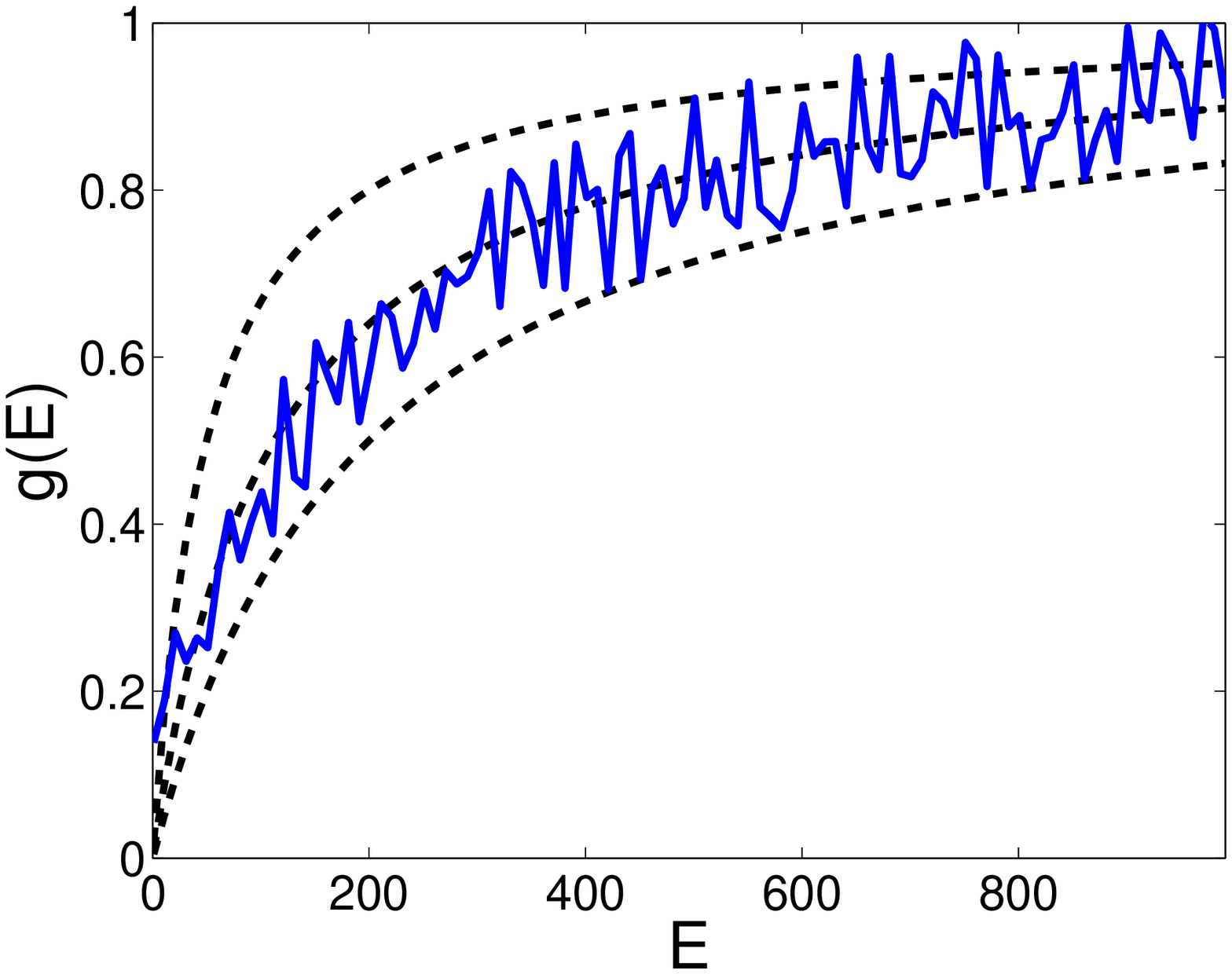}
}

{\footnotesize  
{\bf Fig.11.}
A hypothetical illustration of $g_1(E)$ 
in the case of a complex ``chaotic" barrier.
Such a barrier can be modeled as a network (Fig.1a), 
or it can be characterized using random matrix theory. 
The smooth curves are the transmission $g_0(E;X_2)$ 
of the delta scatterer for 3 different values of $X_2$.     
}
}

\vspace{3mm}

In order to get a degeneracy, a necessary but insufficient 
condition is that the transmission of the two barriers 
is equal (${g_0(E;X_2) = g_1(E)}$). The solution 
of this equation can be determined graphically via Fig.11. 
In fact in most practical applications we can 
assume that our interest is restricted to some 
small energy window such that the smooth $E$ 
dependence of $g_0$ can be neglected. So the 
equation is in fact ${g_0(X_2) = g_1(E)}$. 
For a given $E$ we can find an ${X_2^{(E)}}$ such that this 
equation is satisfied. By playing with $X_1$ we can 
satisfy the $\alpha$ related phase condition 
for having a degeneracy. 
But we still have to satisfy also the $\gamma$ related 
phase condition, which leads to the quantization 
of the energy $E$. Hence the erratic ${X_2^{(E)}}$ 
is sampled. Still it is reasonable to assume that 
the the distribution of the so-obtained $X_2$ values 
is not affected by this random-like sampling. 
We therefore conclude the following relation:   
\be{48}
\mbox{Prob}\Big[X_2 < X_2^{(E)} < X_2+dX_2\big] 
\ \ = \  \
\mbox{Prob}\Big[g_0(X_2) < g_1 < g_0(X_2+dX_2)\big] 
\ee
This implies a simple relation between $\sigma(X_1, X_2)$ 
and the probability function $P(g_1;\bar{g}_1)$
\be{49}
\sigma(X_1,X_2) = \const \times \frac{dg_0(X_2)}{dX_2} P(g_0(X_2)) 
\ee
Thus the problem of finding $\sigma(X_1, X_2)$ has reduced 
to the problem of finding $P(g_1;\bar{g}_1)$.

We can now proceed in three directions: 
{\bf (A)} To determine $P()$ from simple heuristic 
quantum chaos considerations; 
{\bf (B)} To determine $P()$ from formal random  
matrix theory considerations; 
{\bf (C)} To use reverse engineering in order 
to determine what is $P()$ that would give the 
classical result. 
It should be clear that universality can be 
expected only if $\bar{g}_1 \ll 1$.  
In Fig.12 we make a comparison between 
the outcomes of these three procedures 
for $\bar{g}_1 = 0.001$.  In the following 
paragraph we give the details of the calculation. 

\vspace{3mm}

\mpg{
\Cn{
\putgraph[width=0.49\hsize]{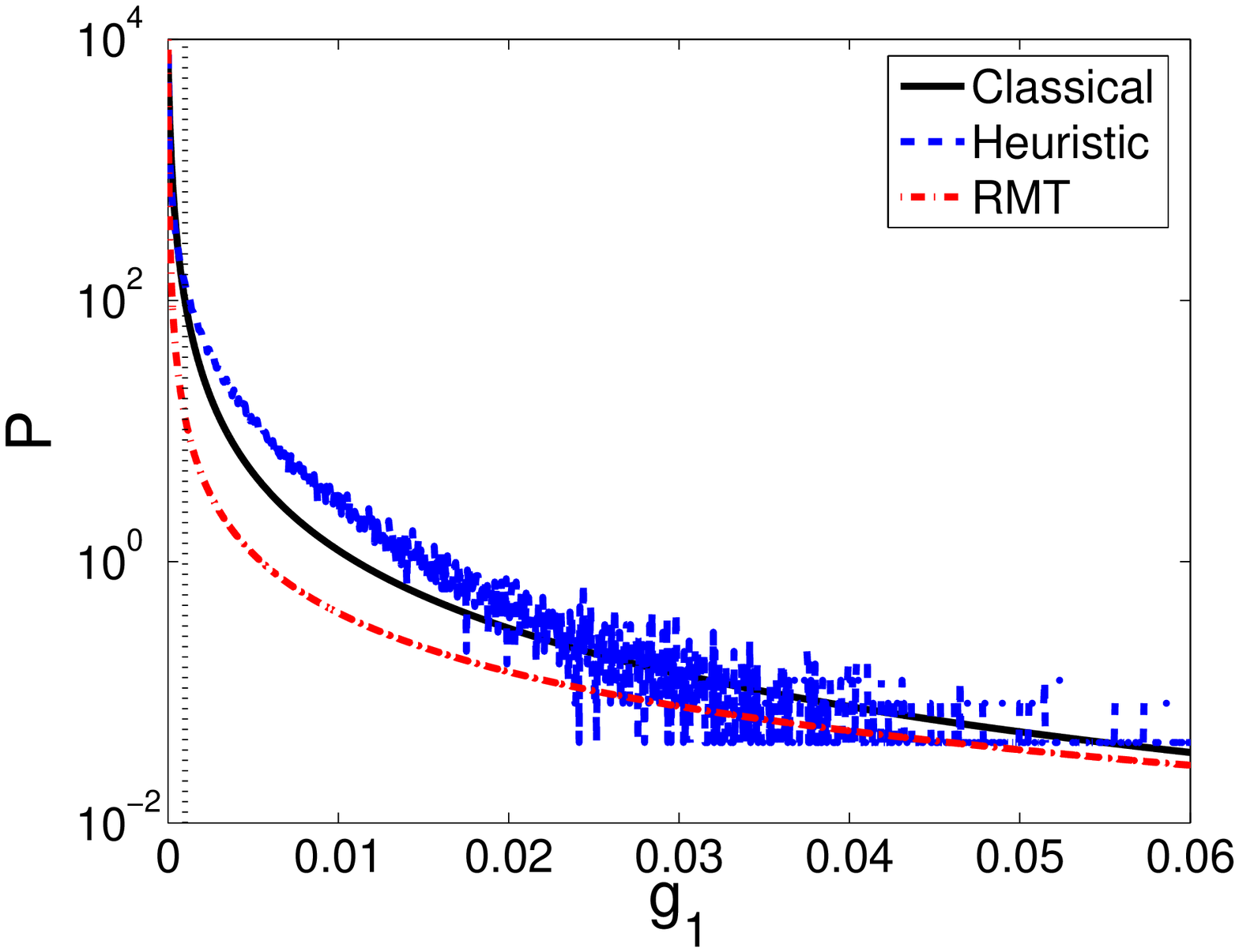}
}

{\footnotesize 
{Fig.12.}
A plot of the distribution $P(g_1;\bar{g}_1)$ according to several different 
expressions. In this calculation we assume that the average  
transmission is ${\bar{g}_1=0.001}$, which is represented in the 
figure by a vertical dashed line. The ``heuristic" result is based on 
sampling of the random variable $g_1=\bar{g}_1\eta_1\eta_2$ where $\eta$ 
is Porter-Thomas distributed. The ``RMT" result is based on Eq.(\ref{e51}).
The ``classical" result is based on Eq.(\ref{e52}). 
}
}

\vspace{3mm}

The heuristic approach is based on the idea that the transmission 
via a chaotic network depends on the amplitudes of the wavefunctions 
at the entrance and exit points. One might expect $g_1=\bar{g}_1\eta_1\eta_2$, 
where $\eta$ has the Porter-Thomas distribution \cite{haake}
${P_{\tbox{GOE}}(\eta) = ({1}/{\sqrt{2\pi\eta}})\eexp{-{\eta}/{2}}}$. 
This leads to the ``heuristic" result in Fig.12. 
In fact this result should not be taken too 
seriously. The formal RMT calculation \cite{BB} 
of the probability distribution $P(g_1;\bar{g}_1)$
leads to the following expressions:
\be{51}
P_{\tbox{RMT}}(g_1;\bar{g}_1) = 
\left\{ 
\amatrix{
({2}/{\pi^2 \bar{g}_1}) \ g_1^{-1/2} & \ \ \mbox{for} \ g_1 \ll (\bar{g}_1)^2 \ll 1 
\cr
({4\bar{g}_1}/{\pi^2}) \ g_1^{-3/2} & \ \ \mbox{for} \ (\bar{g}_1)^2 \ll g_1 \ll 1
}\right.
\ee
The small $g_1$ approximation is universal:   
it merely assumes that the system has time reversal 
symmetry. It has been confirmed \cite{KS} 
that this universal behavior holds also for network 
systems. But for larger values of $g_1$ there are 
deviations that has to do with semiclassical considerations. 
It is therefore in the latter region where one might 
expect quantum-classical correspondence.

The probability distribution $P(g_1;\bar{g}_1)$
that would reproduce the classical 
result Eq.(\ref{e15}) via Eq.(\ref{e49}) is:
\be{52}
P_{\tbox{CL}}(g_1;\bar{g}_1) =
\frac{(1-g^{cl}_1)g^{cl}_1}{(g_1+g^{cl}_1-2g_1g^{cl}_1)^2}
\ee
with ${g_1^{cl} \approx 0.12\bar{g_1}}$. 
In order to compare with the RMT result we note that 
\be{53}
P_{\tbox{CL}}(g_1;\bar{g}_1) \approx 
\left\{ 
\amatrix{
(1/g^{cl}_1) \ (1-2g_1/g_1^{cl}) & \ \ \ \mbox{for} \ g_1 \ll g^{cl}_1 \ll 1 \cr
g^{cl}_1 \ g_1^{-2} & \ \ \ \mbox{for} \ g^{cl}_1 \ll g_1 \ll 1 
}\right.
\ee
We see that in the large $g_1$ region, where one  
might expect quantum-classical correspondence,  
there is no agreement between  $P_{\tbox{CL}}()$ 
and $P_{\tbox{RMT}}()$. 
We suspect that $P_{\tbox{RMT}}()$ cannot 
be trusted there. Otherwise we have to conclude  
that Eq.(\ref{e49}) fails to take into account  
strong correlations in the arrangement of Dirac monopoles.
Either way it seems that RMT alone 
is not enough in order to reproduce the classical result.

\section{The emergence of the classical limit}

With simple minded RMT reasoning we have failed to 
get a quantitative correspondence with the classical result.   
We therefore look for a different way to get 
an estimate for either $\bm{B}_2$ or $\sigma(X_1,X_2)$
in the case of a chaotic network. One obvious 
way is to use the result of Ref.\cite{diabolic} 
regarding the distribution of degeneracies (diabolic points).
The perturbation term which is associated with $X_2$ is 
\be{530}
\mathcal{W} 
= \frac{\partial\mathcal{H}}{\partial X_2}
= \delta(x-X_1)
\ee
and the density of the degeneracies 
should be \cite{diabolic}  
\be{0} 
\sigma(X_1,X_2) 
\ \ = \ \ 
\frac{\pi}{3} \mathsf{g}(E)^2 
\ \mbox{RMS}[\mathcal{F}_{nm}] 
\ \mbox{RMS}[\mathcal{W}_{nm}] 
\ \ \propto \ \  \mbox{RMS}[\mathcal{W}_{nn}]
\ee
where $\mathsf{g}(E)$ is the density of states.
In the first equality it is implicit that 
the root mean square (RMS) of {\em near~diagonal} 
matrix elements should be estimated. 
In fact only  $\mbox{RMS}[\mathcal{W}_{nm}]$ 
is required in order to find the $X_2$ dependence.
For a quantum chaos system with time reversal symmetry 
the variance of the near diagonal elements 
equals half the variance of the diagonal 
elements \cite{agam}, leading to the second expression.

There is a well known semiclassical recipe \cite{mario1,mario2} 
for calculating the variance of the near diagonal 
matrix elements $\mathcal{W}_{nm}$. 
One should find the classical correlation function 
${C(\tau)=\langle \mathcal{W}(t) \mathcal{W}(0) \rangle - \langle \mathcal{W} \rangle^2 }$,  
and then integrate over $\tau$. 
If $\mathcal{W}$ were the current operator 
then $\langle \mathcal{W} \rangle$ would be equal to zero, 
and we could proceed as in section~5. 
But in case of Eq.(\ref{e530}) there is a problem: 
The sign of $\mathcal{W}(t)$ does not fluctuate, 
and it is essential to take into account 
the distribution of the delay times 
inside the network. Therefore there is no obvious 
relation to the transmissions $g_0$ and $g_1$.

An optional possibility is to try to 
evaluate  $\mbox{RMS}[\mathcal{W}_{nn}]$, 
where $\mathcal{W}_{nn}=|\psi_{\tbox{barrier}}|^2$ 
is the ``intensity" of the wavefunction 
at the location of the scatterer. Obviously 
the result depends on both $g_0$ and $g_1$, 
and requires considerations which are at least 
as difficult as estimating universal conductance 
fluctuations. So it seems that we would run 
into the same problems as in the previous section.

Still there is the option to calculate $G^1=\bm{B}_2$ 
from the Green function of the system. 
This has been done in \cite{pmt}:  
Writing the Green function as a sum over trajectories, 
we have expressed $G^1$ as a double sum over paths.
If this double sum is averaged over the energy 
one obtains the diagonal approximation, 
leading to the classical result. At first glance 
the energy averaging is not quite legitimate, 
because the energy is quantized. But one can justify 
this procedure in the case of a ``quantum chaos system".  
We have further supported this claim by the numerical 
analysis of the chaotic network of Fig.1 \cite{pmt}.   
We therefore conclude that for a chaotic network 
the distribution of degeneracies should be in accordance 
with Eq.(\ref{e15}).

\section{Conclusions}

As we translate a scatterer of ``size" $X_2$ 
a distance $\Delta X_1$ along a single mode wire, 
the amount of charge which is pushed is    
\be{0}
Q \ \ = \ \ r(X_2) \times \frac{e}{\pi}k_{\tbox{F}} \times \Delta X_1
\ee
where $k_{\tbox{F}}$ is the Fermi momentum. If the 
scatterer is very ``large" (${X_2\rightarrow\infty}$)
then we expect to have $r(X_2)=1$. 
This expectation is based on the ``snow plow" picture 
that has been explained in the conclusion of section~2.  
This result is also confirmed by the formal BPT based 
calculation in the case of an {\em open} geometry. 
It also can be formally derived for a {\em closed}  
geometry using the ``Dirac chains picture". 
In the latter case the key observation is 
that the $X_1$ distance between contributing 
degeneracies is roughly half the De-Broglie wavelength.
See Eq.(\ref{e34}).

Next we ask what happens to  $r(X_2)$ as $X_2$ becomes smaller. 
In the case of an {\em open} geometry the intuitive naive 
guess, which is based on the ``snow plow" picture,  
turns out to be correct. 
Namely, ${r(X_2)=1-g_0 }$ is simply the reflection 
coefficient: Some of particles are not ``pushed" 
by the scatterer because of its partial transparency.   
In the case of a {\em closed} geometry we have 
shown that the {\em classical} result for $r(X_2)$ is modified: 
now it depends also on the overall transmission of 
the device. See Eq.(\ref{e13}).

It is important to realize that the {\em classical} result 
for $r(X_2)$ is in complete agreement with the common  
sense expectation. Namely, we have ${0 < r(X_2) < 1 }$, 
and the dependence on the ``size" of the 
scatterer is monotonic. 
But once we go to the quantum mechanical analysis 
we have a surprise. The results that we get  
are counter-intuitive. They are most puzzling (Fig.10) 
in the case of the simplest model, in which the ring contains 
only one fixed delta barrier ($V$). 
As we decrease $X_2$ the transported charge $Q$ 
becomes larger(!). Moreover, once $X_2$ becomes 
smaller than $V$, the coefficient $r(X_2)$ 
changes sign. This means that as we push the particles 
``forward" the current is induced ``backwards".

The reason for the failure of our intuition is 
our tendency to regard ``adiabatic transport"  
as a zero order adiabatic approximation, 
while in fact it is based on a first order analysis 
(for a detailed discussion see section~4 of \cite{pms}). 
As a parameter in the system  is changed, 
the induced current can be in either direction.

In order to understand the route towards quantum-classical 
correspondence it is essential to figure out how 
the degeneracies spread out in $\vec{X}$ space. 
As the system becomes more complex, we get for $r(X_2)$   
a result that resembles the classically implied one.  
The resemblance is at best only on a coarse grained scale: 
the quantum result has strong fluctuations.   
These are related to universal conductance fluctuations. 
   
We have made an attempt to deduce from RMT considerations 
the ``chaotic" distribution of the degeneracies, 
and hence the dependence of $r(X_2)$ on $X_2$. 
The quantitative results do not agree. We therefore 
suspect that RMT considerations alone are not enough 
in order to establish quantum-classical correspondence.
Rather we had used \cite{pmt} semiclassical tools 
in order to establish this correspondence.


\vspace{3mm} 

\ack

We thank Nir Davidson for illuminating us regarding 
the feasibility of measuring neutral currents in 
cold atom systems. We also had the pleasure to have 
very helpful discussions with Itamar Sela, 
Tsampikos Kottos, Holger Schanz and Michael Wilkinson.  
This research was supported by 
the Israel Science Foundation (grant No.11/02),
and by a grant from the GIF, the German-Israeli Foundation 
for Scientific Research and Development.



\end{document}